\documentclass[12pt,onecolumn,draftcls,peerreview]{IEEEtran} \newcommand{\figsize}{4.1}
\usepackage{graphicx,psfrag,subfigure}
\usepackage[usenames]{color}
\usepackage[cmex10]{amsmath}
\usepackage{amsfonts,amssymb,latexsym,cite,ifthen,color}
\newboolean{SEP_FIG_CAPS}\setboolean{SEP_FIG_CAPS}{false}

\ifthenelse{\boolean{SEP_FIG_CAPS}}
{
 \newcommand{\putFrag}[4]{\begin{figure}[p]
                            \centering
                            #4
			    \includegraphics[width=#3in]{figures/#1.eps}
            		    \caption{}
     			    \label{fig:#1}
                          \end{figure}
                          \clearpage}
 
 \newcommand{\capFrag}[2]{\noindent Fig.~\ref{fig:#1}. #2 \medskip\\}
 \newcommand{\capTable}[2]{\noindent Tab.~\ref{tab:#1}. #2 \medskip\\}
}
{
 \newcommand{\putFrag}[4]{\begin{figure}[h]
                            \centering
                            #4
			    \includegraphics[width=#3in]{figures/#1.eps}
            		    \caption{#2}
           		    \label{fig:#1}
                          \end{figure} }
 
 \newcommand{\capFrag}[2]{}
 \newcommand{\capTable}[2]{}
}

 \newcommand{\hvec}[1]{\ensuremath{\Hat{\boldsymbol{#1}}}}
    
 \renewcommand{\vec}[1]{\ensuremath{\boldsymbol{#1}}}

 \newcommand{\mc}[1]{\ensuremath{\mathcal{#1}}}

 \newcommand{\Real}{{\mathbb{R}}}
 \newcommand{\Complex}{{\mathbb{C}}}

 \newcommand{\defn}{\triangleq}
 \newcommand{\overall}{_{\text{\sf tot}}}
 \newcommand{\con}{_{\text{\sf con}}}
 \newcommand{\ignore}[1]{}

 \DeclareMathOperator{\E}{E}

 \DeclareMathOperator*{\argmax}{arg\,max}
 \DeclareMathOperator*{\argmin}{argmin}

 \newtheorem{lemma}{Lemma}

 \renewcommand{\eqref}[1]{(\ref{eq:#1})}
 
 \newcommand{\Figref}[1]{Figure~\ref{fig:#1}}
 \newcommand{\figref}[1]{Fig.~\ref{fig:#1}}
 \newcommand{\tabref}[1]{Table~\ref{tab:#1}}
 \newcommand{\secref}[1]{Section~\ref{sec:#1}}

 \newcommand{\lemref}[1]{Lemma~\ref{lem:#1}}

 \newcommand{\ie}{i.e., }
 \newcommand{\eg}{e.g., }
 \newcommand{\mumax}{\mu_{\text{\sf max}}}
 \newcommand{\mumin}{\mu_{\text{\sf min}}}

 \newcounter{comment}[section]
 
 \newcounter{texthead}[section]

\begin{document}
\setlength{\arraycolsep}{0.8mm}
 \title{Joint Scheduling and Resource Allocation in OFDMA Downlink Systems via ACK/NAK Feedback}
	 \author{
	 \IEEEauthorblockN{Rohit Aggarwal,
	 		C.~Emre Koksal, and 
			Philip Schniter} \\
         \IEEEauthorblockA{Dept. of ECE,
	 		   The Ohio State University,
			   Columbus, OH 43210. \\
			   Email: \{aggarwar,koksal,schniter\}@ece.osu.edu} \\
	}
 \date{\today}
 \maketitle

\begin{abstract}
In this paper, we consider the problem of joint scheduling and resource 
allocation in the OFDMA downlink, 
with the goal of maximizing an expected long-term goodput-based utility 
subject to an instantaneous sum-power constraint, 
and where the feedback to the base station 
consists only of ACK/NAKs from recently scheduled users. 
We first establish that the optimal solution is a partially observable 
Markov decision process (POMDP), which is impractical to implement.
In response, we propose a greedy approach to joint scheduling and resource 
allocation that maintains a posterior channel distribution for every user,
and has only polynomial complexity.
For frequency-selective channels with Markov time-variation, we then
outline a recursive method to update the channel posteriors, based on 
the ACK/NAK feedback, that is made computationally efficient through 
the use of particle filtering. 
To gauge the performance of our greedy approach relative to that of the 
optimal POMDP, we derive a POMDP performance upper-bound.
Numerical experiments show that, for slowly fading channels,
the performance of our greedy scheme is relatively close to the upper
bound, and much better than fixed-power random user scheduling (FP-RUS), 
despite its relatively low complexity.

\end{abstract}

\clearpage

\section{Introduction} 				\label{sec:intro}
In the downlink of a wireless orthogonal frequency division multiple access 
(OFDMA) system, the base station (BS) must deliver data to a set of users 
whose channels may vary in both time and frequency. Since bandwidth and power 
resources are limited, data delivery must be carried out efficiently, \eg by 
pairing users with strong subchannels and by distributing power across users 
in the most effective manner. Often, the BS must also adhere to per-user 
quality-of-service (QoS) constraints. Overall, the BS faces the challenging 
problem of jointly scheduling users across subchannels, optimizing their 
modulation-and-coding schemes, and allocating a limited power resource to 
maximize some function of per-user throughputs. 

The OFDMA scheduling-and-resource-allocation problem has been addressed in 
a number of studies that assume the availability of perfect channel state 
information (CSI) at the BS (\eg \cite{Song1,Oct:JSAC:Wong:99, 
Feb:JSAC:Jang:03,Mar:TIT:Willink:97, Jun:TCOM:Hoo:04,Apr:PICA:Wong:07,Jul:PIIT:Seong:06}).
In practice, however, it is difficult for the BS to maintain perfect CSI 
(for all users and all subchannels), since CSI is most easily obtained at 
the user terminals, and the bandwidth available for feedback of CSI to 
the BS is scarce. Hence, practical resource allocation schemes use some form 
of limited feedback \cite{Love:JSAC:08}, such as quantized channel gains.

In this work, we consider the exclusive use of ACK/NAK feedback, as provided 
by the automatic repeat request (ARQ) \cite{bertsekasnet} mechanism present 
in most wireless downlinks. We assume standard ARQ,\footnote{ 
  The approach we develop in this paper could be easily extended to other 
  forms of link-layer feedback, e.g., Type-I and Type-II Hybrid ARQ.
  For simplicity and ease of exposition, however, we consider only
  standard ARQ.} 
where every scheduled user provides the BS with either an acknowledgment 
(ACK), if the most recent data packet has been correctly decoded, or a 
negative acknowledgment (NAK), if not. Although ACK/NAKs do not provide 
direct information about the state of the channel, they do provide 
\emph{relative} information about channel quality that can be used for 
the purpose of transmitter adaptation (e.g., \cite{Dec:TWC:Karmokar:06,Aug:TWC:Aggarwal:09}). 
For example, if an NAK was received for a particular packet, then it is 
likely that the subchannel's signal-to-noise ratio (SNR) was below that 
required to support the transmission rate used for that packet. 
We consider the \emph{exclusive} use ACK/NAK feedback provided by the 
link layer, because this allows us to completely avoid \emph{any additional} 
feedback, such as feedback about quantized channel gains.

There are interesting implications to the use of (quantized) 
\emph{error-rate} feedback (like ACK/NAK) for transmitter adaptation, as 
opposed to quantized channel-state feedback. With error-rate feedback, 
the transmission parameters applied at a given time-slot affect not only 
the throughput for that slot, but also the corresponding feedback, which 
will impact the quality of future transmitter-CSI, and thus future throughput.
For example, if the transmission parameters are chosen to maximize only the 
instantaneous throughput, e.g., by scheduling those users that the BS 
believes are currently best, then little will be learned about the changing 
states of other user channels, implying that future scheduling decisions 
will be compromised. On the other hand, if the BS schedules 
not-recently-scheduled users solely for the purpose of probing their 
channels, then instantaneous throughput will be compromised. Thus, when 
using error-rate feedback, the BS must navigate the classic tradeoff between 
exploitation and exploration~\cite{Jan:MS:Manohan:82}.

In this work, we propose a scheme whereby the BS uses ACK/NAK feedback to maintain a posterior channel distribution for every user and, from these distributions, performs simultaneous user subchannel-scheduling, power-allocation, and rate-selection.
In doing so, the BS aims to maximize an expected, long-term, generic \emph{utility} criterion that is a function of the per-user/channel/rate goodputs.
Our use of a generic utility-based criterion allows us to handle, e.g., sum-capacity maximization, throughput maximization under practical modulation-and-coding schemes, and throughput-based pricing (e.g., \cite{Pricingsurvey, Pricingsurvey2,shenker}), as discussed in the sequel. 
To this end, we exploit our recent work \cite{aggarwal:TSP:2010}, which offers an efficient near-optimal scheme for utility-based OFDMA resource allocation under distributional CSI.
Our use of ACK/NAK-feedback, however, makes our problem considerably more complicated than the one considered in \cite{aggarwal:TSP:2010}.
For example, as we show in the sequel, the optimal solution to our expected long-term utility-maximization problem is a \emph{partially observable Markov decision process} (POMDP) that would involve the solution of many mixed-integer optimization problems during each time-slot. 
Due to the impracticality of the POMDP solution, we instead consider (suboptimal) \emph{greedy} utility-maximization schemes.
As justification for this approach, we first establish that the optimal utility maximization strategy would itself be greedy if the BS had perfect CSI for all user-subchannel combinations. 
Moreover, we establish that the performance of this perfect-CSI (greedy) scheme upper-bounds the optimal ACK/NAK-feedback-based (POMDP) scheme.
We then propose a novel, greedy utility-maximization scheme whose performance is shown (via the upper bound) to be close to optimal. 
Finally, due to the computational demands of tracking the posterior channel distribution for every user, we propose a low-complexity implementation based on particle filtering.

We now describe the relation of our work to the existing 
literature~\cite{haleem,wanglau,holau}.
In \cite{haleem}, a \emph{learning-automata}-based user/rate
scheduling algorithm was proposed to maximize system throughput 
based on ACK/NAK feedback while satisfying per-user throughput constraints.
While \cite{haleem} considered a single channel, we consider joint user/rate 
scheduling and power allocation in a multi-channel OFDMA setting.
In~\cite{wanglau}, a state-space-based approach was taken 
to jointly schedule users/rates and allocate powers 
in downlink OFDMA systems under
slow-fading channels in the presence of ACK/NAK feedback and imperfect 
subchannel-gain estimates at the BS. In particular, assuming 
a discrete channel model, goodput maximization was considered under a target 
maximum packet-error probability constraint and a sum-power constraint 
across all time-slots. Its solution led to a POMDP 
which was solved using a dynamic-program.
While the approach in~\cite{wanglau} is applicable to only goodput 
maximization under discrete-state channels, ours is applicable to generic 
utility maximization problems under continuous-state channels. Furthermore, 
our approach is based on particle filtering and lends itself to 
practical implementation. 
 In~\cite{holau}, 
the user/rate scheduling and power allocation problem in OFDMA systems 
with quasi-static channels and ACK/NAK feedback was formulated as a Markov 
Decision Process and an efficient algorithm was proposed to 
maximize achievable sum-rate while maintaining a target 
packet-error-rate and a sum-power constraint over a finite time-horizon.
Apart from assuming a discrete-state quasi-static channel model, the scope of this work 
was limited by two other assumptions: \emph{i)}
in each time-slot, the BS scheduled only 
one user across all subchannels for data transmission, and \emph{ii)}
all users decoded the broadcasted data-packet and sent ACK/NAK 
feedback to the BS.
In contrast, we consider the scenario 
where multi-user diversity is efficiently exploited by scheduling
different users across different subchannels, and only the
scheduled users report ACK/NAK feedback.
Furthermore, we consider general utility maximization under
continuous-state time-varying channels, and propose a
polynomial-complexity joint scheduling and resource allocation
scheme with provable performance guarantees.

The rest of the paper is organized as follows. In \secref{model}, we 
outline the system model and, in \secref{optimal},
we investigate the optimal scheduling and resource allocation scheme. 
Due to the implementation complexity of the optimal scheme, we propose 
a suboptimal greedy scheme in \secref{greedy} that maintains posterior
channel distributions inferred from the received ACK/NAK feedback. 
In \secref{SSGupdate}, we show how these posteriors can be recursively 
updated via particle filtering. 
Numerical results are presented in \secref{simulations},
and conclusions are stated in \secref{conclusion}.

\section{System Model}					\label{sec:model}

We consider a packetized downlink OFDMA system with a pool of $K$ users.
During each time slot, the BS (\ie ``controller'') transmits packets of data, composed of codewords from a generic signaling scheme, through $N$ OFDMA subchannels (with $N \!\lessgtr\! K$). 
Each packet propagates through a fading channel on the way to its intended mobile user, where
the fading channel is assumed to be time-invariant over the packet duration, but is allowed to vary across packets in a Markovian manner. 
Henceforth, we will use ``time'' when referring to the packet index. 
At each time-instant, the BS must decide---for each subchannel---which 
user to schedule, which modulation-and-coding scheme (MCS) to use, and how 
much power to allocate. 

We assume $M$ choices of MCS, where the MCS index $m \in\{1, \ldots, M\}$ corresponds to a transmission rate of $r_m$ bits per packet and a packet error rate of the form
$\epsilon = a_m e^{-b_m P \gamma}$ under transmit power $P$ and squared subchannel gain (SSG) $\gamma$, where $a_m$ and $b_m$ are constants~\cite{TCOM:wong:09}.
Let $(n,k,m)$ represent the combination of user $k$ and MCS $m$ over subchannel $n$. 
In the sequel, we use $P_{n,k,m}^t$, $\gamma_{n,k}^t$, and $\epsilon_{n,k,m}^t$ to denote---respectively---the power allocated to, the SSG experienced by, and the error rate of the combination $(n,k,m)$ at time $t$.  
Additionally, we denote the scheduling decision by $I_{n,k,m}^t\in\{0,1\}$, where $I_{n,k,m}^t=1$ indicates that user/rate $(k,m)$ was scheduled on subchannel $n$ at time $t$, whereas $I_{n,k,m}^t = 0$ indicates otherwise.
Since we assume that only one user/rate $(k,m)$ can be scheduled on a given subchannel $n$ at a given time $t$, we have the ``subchannel resource'' constraint $\sum_{k,m}I_{n,k,m}^t \leqslant 1$ for all $n,t$. 
We also assume a ``sum-power constraint'' of the form $\sum_{n,k,m}I_{n,k,m}^t\,P_{n,k,m}^t \leqslant X\con$ for all $t$. 

Our goal in scheduling and resource allocation is to maximize an expected long-term utility criterion that is a function of the per-user/rate/subchannel goodputs, i.e., $\E\big\{\sum_{n,k,m,t} U_{n,k,m}(g_{n,k,m}^t)\big\}$. 
Here, $g_{n,k,m}^t$ denotes the goodput contributed by user $k$ with MCS $m$ on subchannel $n$ at time $t$, which can be expanded as
$g_{n,k,m}^t = I_{n,k,m}^t(1-\epsilon_{n,k,m}^t)r_m$.
Meanwhile, $U_{n,k,m}(\cdot)$ is a generic utility function that we assume (for technical reasons) is twice differentiable, strictly-increasing, and concave, with $U_{n,k,m}(0) < \infty$. 
We use $U_{n,k,m}(\cdot)$ to transform goodput into other metrics that are more meaningful from the perspective of quality-of-service (QoS), fairness \cite{TWC:zhang:08}, or pricing (e.g., \cite{Pricingsurvey, Pricingsurvey2,shenker}).
For example, to maximize sum-goodput, one would simply use $U_{n,k,m}(x)=x$.
To enforce fairness across users, one could instead maximize weighted sum-goodput via $U_{n,k,m}(x)= w_k x$, where $\{w_k\}$ are appropriately chosen user-dependent weights.
To maximize sum capacity, i.e.,
$\sum_{n,k}I_{n,k,1}^t \log (1+P_{n,k,1}^t\gamma_{n,k}^t)$, one would
choose $M=a_1 = b_1 = r_1 = 1$ and 
$U_{n,k,1}(x) = \log (1 - \log(1- x))$ for $x\in [0, 1)$. 
To incorporate user-fairness into capacity maximization, one could instead choose 
$U_{n,k,1}(\cdot) = w_{k}\log (1 - \log(1- x))$, where again
$\{w_{k}\}$ are appropriately chosen user-dependent weights \cite{TCOM:wong:09}.

For each time $t$, the BS performs scheduling and resource allocation based on posterior distributions on the SSGs $\{\gamma_{n,k}^t\}$ inferred from previously received ACK/NAK feedback.
In the sequel, we write the ACK/NAK feedback about the packet transmitted to user $k$ across subchannel $n$ at time $t$ by $f_{n,k}^t\in\{1,0,\emptyset\}$, where $1$ indicates an ACK, $0$ indicates a NAK, and $\emptyset$ covers the case that user $k$ was not scheduled on subchannel $n$ at time $t$.
Thus, in the case of an infinite past horizon and a feedback delay of $d\geqslant 1$ packets, the BS would have access to the feedbacks $\{f_{n,k}^\tau~\forall n,k\}_{\tau=-\infty}^{t-d}$ for time-$t$ scheduling.

\section{Optimal Scheduling and Resource Allocation} \label{sec:optimal}

In this section, we describe the optimal solution to the problem of scheduling 
and resource allocation over the finite time-horizon $t\in\{1,\dots,T\}$. 
For this purpose, some additional notation will be useful.
To denote the collection of all time-$t$ scheduling variables $\{I_{n,k,m}^t\}$,
we use $\vec{I}^t\in\{0,1\}^{NKM}$.
To denote the collection of all time-$t$ powers $\{P^t_{n,k,m}\}$, 
we use $\vec{P}^t\in[0,\infty)^{NKM}$.
To denote the collection of all time-$t$ ACK/NAK feedbacks $\{f_{n,k}^t\}$ 
we use $\vec{F}^t\in\{1,0,\emptyset\}^{NK}$, and to denote the collection of all time-$t$ user-$k$ feedbacks we use $\vec{f}_k^t\in\{1,0,\emptyset\}^{N}$.

For time-$t$ scheduling and resource allocation, the controller has access to 
the previous feedback 
$\vec{F}_{-\infty}^{t-d} \defn \{\vec{F}^{-\infty},\ldots,\vec{F}^{t-d}\}$,
scheduling decisions 
$\vec{I}^{t-d}_{-\infty} \defn \{\vec{I}^{-\infty},\ldots,\vec{I}^{t-d}\}$, 
and power allocations 
$\vec{P}^{t-d}_{-\infty} \defn \{\vec{P}^{-\infty},\dots,\vec{P}^{t-d}\}$.
It then uses this knowledge 
to determine the schedule $\vec{I}^t$ and power allocation $\vec{p}^t$ maximizing the expected utility of the current and remaining packets:
\newcommand{\opt}{\textsf{opt}}
\begin{eqnarray}
(\vec{I}^{t,\opt}, \vec{P}^{t,\opt}) 
&=& \argmax_{(\vec{I}^t, \vec{P}^t) \in \mc{X}} \, 
	\E\bigg\{ \sum_{n,k,m} I_{n,k,m}^t U_{n,k,m}\big( 
	(1-a_m e^{-b_m P_{n,k,m}^t \gamma_{n,k}^t})r_{m}\big) 
\nonumber \\ && \mbox{} 
	+ \sum_{\tau = t+1}^T I_{n,k,m}^{\tau,\opt} U_{n,k,m}\big( 
	(1-a_m e^{-b_m P_{n,k,m}^{\tau,\opt} \gamma_{n,k}^\tau})r_{m}\big) 
	\bigg| \vec{F}_{-\infty}^{t-d}, 
	\vec{I}_{-\infty}^{t-d}, \vec{P}_{-\infty}^{t-d}\bigg\}, 
							\label{eq:pomdp}
\end{eqnarray}
where the domain of $\vec{I}^t$ is 
$\mc{I} \defn \{\vec{I} \in \{0, 1\}^{NKM}: \sum_{k,m}I_{n,k,m}\leqslant 1 ~\forall n\}$, the domain of $\vec{P}^t$ is 
$\mc{P} \defn [0,\infty)^{NKM}$, 
and $\mc{X} \defn \{(\vec{I}, \vec{P}) \in \mc{I}\times\mc{P}:
\sum_{n,k,m} I_{n,k,m} \, P_{n,k,m} \leqslant X\con\}$. 
The expectation in \eqref{pomdp} is jointly over the squared subchannel gains
(SSGs) $\{\gamma_{n,k}^\tau:\tau = t,\dots,T,  \forall n, \forall k \}$.
Using the abbreviations
$\tilde{U}_{n,k,m}^t(I_{n,k,m}, P_{n,k,m}) \defn
I_{n,k,m} U_{n,k,m}\big( (1-a_m e^{-b_m P_{n,k,m} \gamma_{n,k}^t})r_{m}\big)$ 
and $\vec{\mathbb{F}}_{-\infty}^{t-d} \defn
\{\vec{F}_{-\infty}^{t-d}, \vec{I}_{-\infty}^{t-d}, \vec{P}_{-\infty}^{t-d}\}$,
the optimal expected utility over the remaining packets 
$\{t, \ldots, T\}$ can be written (for $t\geqslant 0$) as
\begin{equation}
U^{t,\opt}_{\text{\sf tot}}(
\vec{\mathbb{F}}_{-\infty}^{t-d}) 
\defn \E\bigg\{\sum_{\tau = t}^T \sum_{n, k, m}
	\tilde{U}_{n,k,m}^\tau\big(I_{n,k,m}^{\tau,\opt},
	P_{n,k,m}^{\tau,\opt}\big)\,\bigg| \,
	\vec{\mathbb{F}}_{-\infty}^{t-d}\bigg\}.
\end{equation}
For a unit-delay\footnote{
  For the $d > 1$ case, the Bellman equation is more complicated, and so
  we omit it for brevity.} 
system (i.e. $d = 1$), the following Bellman equation \cite{bertsekas} 
specifies the corresponding finite-horizon dynamic program:
\begin{eqnarray}
U^{t,\opt}_{\text{\sf tot}}(
\vec{\mathbb{F}}_{-\infty}^{t-1})
&=&   \max_{(\vec{I}^t, \vec{P}^t) \in \mc{X}}  \bigg[
	\E \Big\{ \sum_{n,k,m}
	\tilde{U}^t_{n,k,m}\big( I_{n,k,m}^{t},P_{n,k,m}^{t}\big) \,\Big|\,
	\vec{\mathbb{F}}_{-\infty}^{t-1}\Big\}
\nonumber \\ & & \mbox{} 
	+ \E\Big\{U^{t+1,\opt}_{\text{\sf tot}}\big(
	\vec{\mathbb{F}}_{-\infty}^{t-1} \cup
	\{\vec{F}^t,\vec{I}^t,\vec{P}^t\}
	\big) \,\Big|\,\vec{\mathbb{F}}_{-\infty}^{t-1}\Big\} \bigg], 
						\label{eq:bellman}
\end{eqnarray}
where the second expectation is over the feedbacks $\vec{F}^t$. 
The solution obtained by solving \eqref{pomdp} is typically referred to
as a partially observable Markov decision process (POMDP) 
\cite{Jan:MS:Manohan:82}.

The definition of $\mc{P}$ implies that  
the controller has an uncountably infinite number of possible actions. 
Although this could be circumvented (at the expense of 
performance) by restricting the powers $P_{n,k,m}^t$ to come from a finite 
set, the problem would remain very complex due to the continuous-state 
nature of the SSGs $\gamma_{n,k}^t$.
While these SSGs could then be quantized (causing additional performance
loss), the problem would still remain computationally intensive,
since POMDPs (even with finite states and actions) 
are PSPACE-complete, i.e., they require both complexity and memory that 
grow exponentially with the horizon $T$ \cite{OR:Papadimitriou:87}.
To see why, notice from \eqref{bellman} that the solution of the problem 
at every time $t$ depends on the optimal solution at times up to $t\!-\!1$. 
Because both terms on the right side of \eqref{bellman} are dependent 
on $(\vec{I}^t,\vec{P}^t)$, however, the solution of the problem at time $t$ 
also depends on the solution of the problem at time $t+1$, which in turn 
depends on the solution of the problem at time $t+2$, and so on.
In conclusion, the optimal controller is not practical to implement,
even under power/SSG quantization.

Consequently, we will turn our attention to (sub-optimal) 
\emph{greedy} strategies, i.e., those that do not consider the effect 
of current actions on future utilities. 
To better understand their performance relative to that of the optimal POMDP, 
we derive an upper bound on POMDP performance.

\subsection{The ``Causal Global Genie'' Upper Bound} \label{sec:causalgenie}

Our POMDP-performance upper-bound, which we will refer to as the ``causal
global genie'' (CGG), is based on the presumption of \emph{perfect} 
error-rate feedback of \emph{all} previous user/subchannel 
combinations, i.e., 
$\{\epsilon_{n,k,m}^\tau\, \forall n, k, \tau \leqslant t-d\}$. 
For comparison, the ACK/NAK feedback available to the POMDP is a form 
of \emph{degraded} error-rate feedback on \emph{previously scheduled} 
user/subchannel combinations.
Since, given knowledge of $\epsilon_{n,k,m}^\tau$ and $P_{n,k,m}^\tau$
for any rate index $m$, 
the SSG $\gamma_{n,k}^\tau$ can be obtained by simply 
inverting the error-rate expression 
$\epsilon_{n,k,m}^\tau = a_m e^{-b_m P_{n,k,m}^\tau \gamma_{n,k}^\tau}$,
our genie-aided bound is based, equivalently, on perfect feedback of 
all previous SSGs $\{\gamma_{n,k}^\tau\, \forall n, k, \tau \leqslant t-d\}$. 
In the sequel, we use
$\vec{\gamma}^t\in[0,\infty)^{NK}$ to denote the collection of all time-$t$ SSGs $\{\gamma_{n,k}^t\,\forall k,n\}$, and we define
$\vec{\gamma}^{t-d}_{-\infty} \defn \{\vec{\gamma}^{-\infty},\dots,\vec{\gamma}^{t-d}\}$.

We characterize the CGG as ``global'' since it uses feedback from \emph{all} 
user/subchannel combinations, not just the previously scheduled ones.
Although a tighter bound might result if the (perfect) error-rate feedback 
was restricted to only previously scheduled user/subchannel pairs, the bounding 
solution would remain a POMDP with an uncountable number of state-action 
pairs, making it impractical to evaluate.
Evaluating the performance of the CGG, however, is straightforward 
since---under CGG feedback---optimal scheduling and resource 
maximization can be performed greedily. 
To see why, notice that, for any scheduling time $t \geqslant 0$, the CGG 
scheme allocates resources according to the following mixed-integer 
optimization problem:
\begin{eqnarray}
(\vec{I}^{t, \text{\sf cgg}}, \vec{P}^{t, \text{\sf cgg}})
&=& \argmax_{(\vec{I}^t, \vec{P}^t) 
\in \mc{X}} \, \sum_{n,k,m} \E\bigg\{ 
\tilde{U}_{n,k,m}^t\big( I_{n,k,m}^t, P_{n,k,m}^t\big) 
\nonumber \\ && \mbox{} 
+ \sum_{\tau = t+1}^T 
\tilde{U}^\tau_{n,k,m}\big( I_{n,k,m}^{\tau,\text{\sf cgg}}, P_{n,k,m}^{\tau, 
\text{\sf cgg}}\big) \Big| \vec{I}_{-\infty}^{t-d}, \vec{P}_{-\infty}^{t-d}, 
\vec{\gamma}_{-\infty}^{t-d}\bigg\}.
\end{eqnarray}
Since the choice of $\big\{(\vec{I}^{t+1,\text{\sf cgg}}, 
\vec{P}^{t+1,\text{\sf cgg}}), \ldots, (\vec{I}^{T,\text{\sf cgg}}, 
\vec{P}^{T,\text{\sf cgg}})\big\}$ does not depend on the choice of 
$(\vec{I}^{t,\text{\sf cgg}}, \vec{P}^{t,\text{\sf cgg}})$, 
the previous optimization problem simplifies to 
\begin{eqnarray}
(\vec{I}^{t, \text{\sf cgg}}, \vec{P}^{t, \text{\sf cgg}}) 
&=& \argmax_{(\vec{I}^t, \vec{P}^t) \in \mc{X}} \, \sum_{n,k,m} \E 
\Big\{ \tilde{U}^t_{n,k,m}\big( I_{n,k,m}^t, P_{n,k,m}^t \big) 
\Big| \vec{\gamma}_{-\infty}^{t-d}\Big\}. \label{eq:CGGeq}
\end{eqnarray}

In the following lemma, we formally establish that the utility achieved by 
the CGG upper-bounds that achieved by the optimal POMDP controller with 
ACK/NAK feedback. 
\begin{lemma}                   \label{lem:genie_bnd}
Given arbitrary past allocations $(\vec{I}_{-\infty}^{t-d}, 
\vec{P}_{-\infty}^{t-d})$,
and the corresponding ACK/NAKs $\vec{F}_{-\infty}^{t-d}$, 
the expected total utility for optimal resource allocation under the
latter feedback is no higher than the expected total utility 
under CGG feedback, i.e.,
\begin{eqnarray}
\sum_{n, k, m}\sum_{\tau = t}^T \E\Big\{ \tilde{U}_{n,k,m}^\tau 
\big(I_{n,k,m}^{\tau,\opt},P_{n,k,m}^{\tau,\opt}\big)\,\Big|\, 
\vec{\mathbb{F}}_{-\infty}^{t-d}\Big\}
&\leqslant& \sum_{n, k, m}\sum_{\tau = t}^T \E\Big\{
\tilde{U}_{n,k,m}^\tau \big(I_{n,k,m}^{\tau,\text{\sf cgg}},
P_{n,k,m}^{\tau, \text{\sf cgg}}\big)\,\Big| \,
\vec{\mathbb{F}}_{-\infty}^{t-d}\Big\}. \label{eq:genie_bnd}
\end{eqnarray}
\end{lemma}

The proof of the above lemma follows the same steps as the proof 
of~\cite[Lemma $1$]{Aug:TWC:Aggarwal:09}, which is omitted here
to save space. 
In the next section, we detail the greedy scheduling and resource 
allocation problem and propose a near-optimal solution.

\section{Greedy Scheduling and Resource Allocation} \label{sec:greedy}

The greedy scheduling and resource allocation (GSRA) problem is defined
as follows.
\begin{eqnarray}
\textrm{GSRA} &\defn&\max_{\substack{\\[0.2 mm]\scriptstyle \vec{I}^t \in \mathcal{I}\\[0.8 mm]
\scriptstyle \vec{P}^t \in \mc{P}}} \, 
\sum_{n=1}^N \sum_{k=1}^K \sum_{m=1}^M I_{n,k,m}^t \E\Big\{ 
U_{n,k,m}\big( (1-a_m e^{-b_m P_{n,k,m}^t \gamma_{n,k}^t})r_{m}\big) 
\Big| \vec{\mathbb{F}}_{-\infty}^{t-d}\Big\} \nonumber \\ 
&&\mathrm{s.t.}~~~ \sum_{n,k,m} I_{n,k,m}^t \, P_{n,k,m}^t \leqslant X\con.
\label{eq:greedyproblem}
\end{eqnarray}
Note that, in contrast to the $T$-horizon objective \eqref{pomdp},
the greedy objective \eqref{greedyproblem} does not consider the 
effect of $(\vec{I}^t, \vec{P}^t)$ on future utility. 
As stated earlier, we allow $U_{n,k,m}(\cdot)$ to be any real-valued
function that is twice differentiable, strictly-increasing, and concave,
with $U_{n,k,m}(0) < \infty$. 
Therefore, $U'_{n,k,m}(\cdot) > 0$ and $U''_{n,k,m}(\cdot) \leqslant 0$, 
using $'$ to denote the derivative. 

Since it involves both discrete $(\vec{I}^t)$ and continuous $(\vec{P}^t)$ 
optimization variables, the GSRA problem \eqref{greedyproblem} is a
mixed-integer optimization problem. 
Such problems are generally NP-hard, meaning that polynomial-complexity 
solutions do not exist. 
Thus, in \secref{proposed}, we propose a \emph{near}-optimal algorithm 
for \eqref{greedyproblem} with polynomial complexity.
To better explain that scheme, we first describe, in \secref{brute}, 
a ``brute force'' optimal solution whose complexity grows exponentially 
in $N$, the number of subchannels.

\subsection{Brute-Force Algorithm} \label{sec:brute}

The brute-force approach considers all possibilities of $\vec{I}^t\in\mc{I}$,
each with the corresponding optimal power allocation.
Supposing that $\vec{I}^t=\vec{I}$, the optimal power allocation
can be found by solving the convex optimization problem 
\begin{eqnarray}
&&\max_{\substack{\\[0.2 mm] \scriptstyle \vec{P}\in\mc{P}}}\, 
\sum_{n=1}^N \sum_{k=1}^K \sum_{m=1}^M  I_{n,k,m} \E\Big\{ 
U_{n,k,m}\big((1-a_m e^{-b_m P_{n,k,m} \gamma_{n,k}^t})r_{m}\big) \Big| 
\vec{\mathbb{F}}_{-\infty}^{t-d}\Big\} \nonumber \\ 
&&\mathrm{s.t.}~~~ \sum_{n,k,m} I_{n,k,m}\,P_{n,k,m} \leqslant X\con.  
\label{eq:newproblem} 
\end{eqnarray}
To proceed, we identify the Lagrangian associated with \eqref{newproblem} as 
\begin{eqnarray}
L_{\vec{I}}^t(\mu, \vec{P}) 
&=& \Big(\sum_{n,k,m} I_{n,k,m}\,P_{n,k,m} - X\con\Big)\mu 
\nonumber \\ && \mbox{} 
- \sum_{n,k,m} \E\Big\{I_{n,k,m} U_{n,k,m}\big((1-a_m e^{-b_m P_{n,k,m} \gamma_{n,k}^t})r_{m}\big) \Big| \vec{\mathbb{F}}_{-\infty}^{t-d} \Big\},
\label{eq:lagrangian} 
\end{eqnarray}
which yields the corresponding dual problem 
\begin{eqnarray}
\max_{\mu \geqslant 0} \min_{\vec{P} \in \mc{P}} L_{\vec{I}}^t(\mu, \vec{P})
&=& \max_{\mu \geqslant 0} L_{\vec{I}}^t(\mu, \vec{P}^{*}(\mu)) = L_{\vec{I}}^t(\mu_{\vec{I}}^*, \vec{P}^{*}(\mu^*_{\vec{I}})),
\label{eq:dual}
\end{eqnarray}
where $\mu^*_{\vec{I}}$ and $\vec{P}^{*}(\mu^*_{\vec{I}})$ denote the
optimal Lagrange multiplier and power allocation, respectively.

A detailed solution to \eqref{dual} is given in~\cite{aggarwal:TSP:2010}, 
and so we describe only the main points here.
First, for a given value of the Lagrange multiplier $\mu$, it 
has been shown that the optimal powers equal
\begin{eqnarray}
P_{n,k,m}^{*}(\mu) &=&
{\begin{cases} \tilde{P}_{n,k,m}(\mu) & \textrm{if}~ 0 \leqslant \mu 
\leqslant a_m b_m r_m U'_{n,k,m}\big((1-a_m)r_m\big) \E\big\{\gamma_{n,k}^t \big| 
\vec{\mathbb{F}}_{-\infty}^{t-d} \big\} \\
0 & \textrm{otherwise}, \end{cases}}
\label{eq:P_n^*}
\end{eqnarray}
where $\tilde{P}_{n,k,m}(\mu)$ is defined as the (unique) solution to
\begin{eqnarray}
\mu &=& a_m b_m r_m \E\big\{ U'_{n,k,m}\big((1-a_m 
e^{-b_m \tilde{P}_{n,k,m}(\mu) \gamma_{n,k}^t})r_{m}\big) \gamma_{n,k}^t 
e^{-b_m \tilde{P}_{n,k,m}(\mu) \gamma_{n,k}^t} \big| 
\vec{\mathbb{F}}_{-\infty}^{t-d} \big\}.\label{eq:gntildeb} 
\end{eqnarray}
Then, for a given $\vec{I}$, the optimal value of $\mu$ (i.e., 
$\mu_{\vec{I}}^*$) obeys 
$\mu_{\vec{I}}^* \in [\mumin, \mumax] \subset (0, \infty)$, where
\begin{eqnarray}
\mumin &=& \min_{n,k,m} 
a_m b_m r_m \E\big\{U_{n,k,m}'\big((1-a_m e^{-b_m X\con 
\gamma_{n,k}^t})r_m\big)\gamma_{n,k}^t e^{-b_m X\con \gamma_{n,k}^t}
\big| \vec{\mathbb{F}}_{-\infty}^{t-d} \big\}, \label{eq:mumin} \\
\mumax &=& \max_{n,k,m} a_m b_m r_m U_{n,k,m}'\big((1-a_m)r_m\big) 
\E\big\{\gamma_{n,k}^t \big| \vec{\mathbb{F}}_{-\infty}^{t-d} \big\}, \label{eq:mumax}
\end{eqnarray}
and satisfies $\sum_{n,k,m} I_{n,k,m}\, P_{n,k,m}^{*}(\mu_{\vec{I}}^*) = X\con$.
 
Based on \eqref{P_n^*}-\eqref{mumax}, Table~\ref{table1} details 
the brute-force steps for a given $\vec{I}$. 
In the end, for a specified tolerance $\kappa$, these steps find
$\underline{\mu}$ and $\bar{\mu}$ such that $\mu^*_{\vec{I}} \in 
[\underline{\mu}, \bar{\mu}]$ and $\bar{\mu} - \underline{\mu} < \kappa$.
Using an approximation of $\mu_{\vec{I}}^*$ that lies in $[\underline{\mu}, 
\bar{\mu}]$, the corresponding utility is guaranteed to be no less than 
$\kappa X\con$ from the optimal (for the given $\vec{I}$).
Therefore, by adjusting $\kappa$, one can achieve a performance arbitrarily
close to the optimum. 
Since $|\mc{I}|=(KM+1)^N$ values of $\vec{I}$ must be considered,
the total complexity of the brute-force approach---in terms of the number 
of times \eqref{gntildeb} must be solved---can be shown to be 
\begin{eqnarray}\textstyle
\big\lceil\log_2 (\frac{\mu_{\text{\sf max}} 
- \mu_{\text{\sf min}}}{\kappa})\big\rceil \times (KM+1)^{N-1} NKM,
\label{eq:complexitybrute}
\end{eqnarray}
which grows exponentially with $N$.

\subsection{Proposed Algorithm}	\label{sec:proposed}

We propose to attack the mixed-integer GSRA problem \eqref{greedyproblem} 
using the well known \emph{Lagrangian relaxation} approach \cite{bertsekas}.
In doing so, we relax the domain of the scheduling variables $I_{n,k,m}^t$ 
from the set $\{0,1\}$ to the interval $[0,1]$,
allowing the application of low-complexity dual optimization techniques.
Although the solution to the relaxed problem does not necessarily coincide
with that of the original greedy problem \eqref{greedyproblem}, we
establish in the sequel that the corresponding performance loss is very 
small, and in some cases zero.

The relaxed version of the greedy problem \eqref{greedyproblem} is
\begin{eqnarray}
\text{rGSRA} &\defn& \max_{\substack{\\[0.2 mm]\scriptstyle \vec{I}^t \in 
\mathcal{I}_c\\[0.8 mm] \scriptstyle \vec{P}^t \in \mc{P}}} \, 
\sum_{n=1}^N \sum_{k=1}^K \sum_{m=1}^M I_{n,k,m}^t \E\Big\{ 
U_{n,k,m}\big( (1-a_m e^{-b_m P_{n,k,m}^t \gamma_{n,k}})r_{m}\big) 
\Big| \mathbb{F}^{t-d}_{-\infty}\Big\} \nonumber \\ 
&&\mathrm{s.t.}~~~ \sum_{n,k,m} I_{n,k,m}^t \, P_{n,k,m}^t \leqslant X\con,
\label{eq:relaxedproblem}
\end{eqnarray}
where $\mc{I}_c \defn \big\{\vec{I} \in [0,1]^{NKM}: \sum_{k,m}I_{n,k,m} \leqslant 1 ~\forall n\big\}$. 
Although \eqref{relaxedproblem} is a non-convex optimization problem due 
to non-convex constraints, it can be converted into a convex optimization 
problem by using the new set of variables $(\vec{I}^t, \vec{x}^t)$, 
where $x_{n,k,m}^t \defn I_{n,k,m}^t\,P_{n,k,m}^t$.
In this case, we have
\begin{eqnarray}
\text{rGSRA} = \min_{\substack{\\[0.2 mm]\scriptstyle \vec{x}^t \succeq 0\\[0.8 mm]
	\scriptstyle \vec{I}^t \in \mathcal{I}_c}} 
	\sum_{n,k,m} I_{n,k,m}^t\,B_{n,k,m}^t(I_{n,k,m}^t,x_{n,k,m}^t)
	~~~\text{s.t.}~\sum_{n,k,m} x^t_{n,k,m} \leqslant X\con,
\label{eq:CSRAP}
\end{eqnarray}
where $\vec{x}^t\in\Real^{NKM}$ denotes the collection of all time-$t$ 
variables $\{x_{n,k,m}^t\}$,
$\vec{x}^t \succeq 0$ denotes element-wise non-negativity, 
and $B_{n,k,m}^t(\cdot,\cdot)$ is defined as
\begin{eqnarray}
B_{n,k,m}^t(y_1,y_2) 
&\defn& {\begin{cases}
	- \E\Big\{ U_{n,k,m}\big((1-a_m e^{-b_m \gamma_{n,k}^t \, y_2
	/y_1}) r_{m}\big) \Big| \mathbb{F}^{t-d}_{-\infty}\Big\} 
	& \textrm{if}~ y_1 \neq 0 \\
0 & \textrm{otherwise}. \label{eq:defF}
\end{cases}}
\end{eqnarray}
The modified problem \eqref{CSRAP} is a convex optimization problem and can be
solved using a dual optimization approach with zero duality gap. 
In particular, the dual problem can be written as
\begin{eqnarray}
\lefteqn{
\max_{\substack{\\[0.2 mm]\scriptstyle \mu \geqslant 0}}
\min_{\substack{\\[0.2 mm]\scriptstyle \vec{x}^t\succeq 0 \\[0.8 mm]
	\scriptstyle \vec{I}^t\in \mc{I}_c}}
	L(\mu,\vec{I}^t,\vec{x}^t)
= \max_{\substack{\\[0.2 mm]\scriptstyle \mu \geqslant 0}}
	\min_{\substack{\\[0.2 mm]\scriptstyle \vec{I}^t\in \mc{I}_c}}
	L(\mu,\vec{I}^t,\vec{x}^{t,*}(\mu, \vec{I}^t))} \nonumber \\
&=& \max_{\substack{\\[0.2 mm]\scriptstyle \mu \geqslant 0}}
	L(\mu,\vec{I}^{t,*}(\mu),\vec{x}^{t,*}(\mu, \vec{I}^{t,*}(\mu))) 
= L(\mu^*,\vec{I}^{t,*}(\mu^*),\vec{x}^{t,*}(\mu^*, \vec{I}^{t,*}(\mu^*))), \label{eq:cdual} 
\end{eqnarray}
where
\begin{equation}
L(\mu,\vec{I}^t,\vec{x}^t) 
\defn \sum_{n,k,m}  I_{n,k,m}^t\,B_{n,k,m}^t(I_{n,k,m}^t,
x_{n,k,m}^t) + \Big(\sum_{n,k,m}x_{n,k,m}^t - X\con\Big) \mu,  \label{eq:clagrange}
\end{equation}
where $\vec{x}^{*}(\mu, \vec{I})$ is the optimal $\vec{x}$ for a given $(\mu, 
\vec{I})$, 
where $\vec{I}^*(\mu)$ denotes the optimal 
$\vec{I}\in \mc{I}_c$ for a given $\mu$, 
and where $\mu^*$ denotes the optimal $\mu\geqslant 0$.

A detailed solution to this problem was given in 
\cite{aggarwal:TSP:2010}, and so we describe only the main points here.
For given values of $\mu$ and $\vec{I}^t$, we have
$x_{n,k,m}^{t,*}(\mu, \vec{I}^t) = I_{n,k,m}^t \, P_{n,k,m}^{t,*}(\mu)$, where
\begin{eqnarray}
P_{n,k,m}^{t,*}(\mu) &=&
{\begin{cases} \tilde{P}_{n,k,m}^t(\mu) & \textrm{if}~ 0 \leqslant \mu 
\leqslant a_m b_m r_m U'_{n,k,m}\big((1-a_m)r_m\big) \E\big\{\gamma_{n,k}^t
\big| \mathbb{F}^{t-d}_{-\infty} \big\} \\
0 & \textrm{otherwise}, \end{cases}}
\label{eq:P_n^t*}
\end{eqnarray}
and where $\tilde{P}_{n,k,m}^t(\mu)$ is defined as the (unique) solution to
\begin{eqnarray}
\mu &=& a_m b_m r_m \E\big\{ U'_{n,k,m}\big((1-a_m 
e^{-b_m \tilde{P}^t_{n,k,m}(\mu) \gamma_{n,k}^t})r_{m}\big) \gamma_{n,k}^t 
e^{-b_m \tilde{P}^t_{n,k,m}(\mu) \gamma_{n,k}^t} 
\big| \mathbb{F}^{t-d}_{-\infty} \big\}.\label{eq:gntilde} 
\end{eqnarray}
To give equations that govern $\vec{I}^{t,*}(\mu)$ for a given $\mu$, we first define 
\begin{align}
V_{n,k,m}^t(\mu,P_{n,k,m}^{t,*}(\mu)) 
\defn - \E\Big\{ U_{n,k,m}\big((1 - 
a_m e^{-b_m P_{n,k,m}^{t,*}(\mu) \gamma_{n,k}})r_{m} 
\big) \big| \mathbb{F}^{t-d}_{-\infty} \Big\} + \mu P_{n,k,m}^*(\mu)
\label{eq:V} \\
S_n^t(\mu) \defn \Big\{(k,m) = \argmin_{(k',m')} V_{n,k',m'}^t(\mu, 
P_{n,k',m'}^{t,*}(\mu)): V_{n,k,m}^t(\mu, P_{n,k,m}^{t,*}(\mu)) \leqslant 0\Big\}. \label{eq:Snmu}
\end{align}
If $S_n^t(\mu)$ is a null or a singleton set, then the optimal schedule on 
subchannel $n$ is given by
\begin{eqnarray}
I_{n,k,m}^{t,*}(\mu) = \begin{cases} 1 & (k,m) \in S_n^t(\mu)\\
0 & \text{otherwise}. \end{cases} \label{eq:I}
\end{eqnarray}
However, if $S_n^t(\mu)$ has cardinality greater than one, then multiple 
$(k,m)$ combinations can be scheduled simultaneously while achieving the 
optimal value of the Lagrangian. 
In particular, if
$S_n^t(\mu) = \{(k_1(n),m_1(n)), \ldots, 
(k_{|S_n^t(\mu)|}(n),m_{|S_n^t(\mu)|}(n))\}$,
then 
\begin{eqnarray}
I_{n,k,m}^{t,*}(\mu) = {\begin{cases} I_{n,k_i(n),m_i(n)} & \text{if}~ (k,m) = (k_i(n),m_i(n)) 
~\textrm{for some}~ i \in \{1, \ldots, |S_n^t(\mu)|\} \\
0 & \text{otherwise}, \end{cases}} \label{eq:I*odd}
\end{eqnarray}
where the vector $\big[I_{n,k_1(n),m_1(n)}, \ldots, 
I_{n,k_{|S^t_n(\mu)|}(n),m_{|S^t_n(\mu)|}(n)}\big]$ 
lies anywhere in the unit-$(|S_n^t(\mu)|\!-\!1)$ simplex, 
i.e., it lives within the region $[0,1]^{|S_n^t(\mu)|}$ and satisfies
$\sum_{i=1}^{|S^t_n(\mu)|} I_{n,k_i(n),m_i(n)} = 1$. 
Finally, the optimal Lagrange multiplier $\mu$ (i.e., $\mu^*$) is such that 
$\mu^* \in [\mumin, \mumax] \subset (0, \infty)$ and 
\begin{equation}
\sum_{n,k,m} I_{n,k,m}^{t,*}(\mu^*)\, P_{n,k,m}^{t,*}(\mu^*) = X\con, 
\end{equation}
where $\mumin$ and $\mumax$ were given in 
\eqref{mumin} and \eqref{mumax}, respectively. 

For several fixed values of $\mu$, the proposed algorithm minimizes the 
relaxed Lagrangian \eqref{clagrange} over $(\vec{I}^t,\vec{x}^t)$
(or, equivalently, over $(\vec{I}^t,\vec{P}^t)$) to obtain candidate 
solutions for the original greedy problem \eqref{greedyproblem}. 
If, for a given $\mu$, $|S_n^t(\mu)| \leqslant 1$ for all $n$ 
(i.e., the candidate employs at most one user/MCS per subchannel),
then the candidate solution is admissible for the non-relaxed problem, 
and thus retained by the proposed algorithm.
If, on the other hand, $|S_n^t(\mu)| > 1$ for some $n$ 
(i.e., the candidate employs more than one user/MCS on some subchannels), 
then the proposed algorithm transforms the candidate
into an admissible solution as follows:
\begin{eqnarray}
I^{t,\text{\sf pro}}_{n,k,m}(\mu) = \begin{cases} 1 & (k,m) = \argmin_{(k',m') \in 
S_n^t(\mu)} P_{n,k',m'}^{t,*}(\mu) \\
0 & \text{otherwise}. \end{cases} \label{eq:I*}
\end{eqnarray}
The following lemma then states an important property of these fixed-$\mu$ admissible solutions.
\begin{lemma} \label{lem:lem3}
For any given value of $\mu$, let the power allocation 
$\vec{P}^{t,*}(\mu)$ be given by \eqref{P_n^t*},
let the user-MCS allocation $\vec{I}^{t,\text{\sf pro}}(\mu)$ be given by \eqref{I*}, and let the total power allocation be defined as 
$X\overall^{t,\text{\sf pro}}(\mu) \defn 
\sum_{n,k,m}I_{n,k,m}^{t,\text{\sf pro}}(\mu)\,P_{n,k,m}^{t,*}(\mu)$.
Then, $X\overall^{t,\text{\sf pro}}(\mu)$ is monotonically decreasing in $\mu$.
\end{lemma}

\lemref{lem3} (see \cite{aggarwal:TSP:2010} for a proof) 
implies that the optimal value of the Lagrange multiplier $\mu$
(i.e., $\mu^*$) is the one that achieves the power constraint 
$X\overall^{t,\text{\sf pro}}(\mu)=X\con$.
To find this $\mu^*$, the proposed algorithm performs a 
bisection search over $\mu \in [\mumin, \mumax]$ that refines the
search interval $[\underline{\mu}, \bar{\mu}]$ until
$\bar{\mu}-\underline{\mu} < \kappa$, 
where $\kappa$ is a user-defined tolerance.
Then, between the two schedules
$\vec{I} \in \{\vec{I}^{t,\text{\sf pro}}(\underline{\mu}), \vec{I}^{t,\text{\sf pro}}(\bar{\mu})\}$,
it chooses the one that maximizes utility, reminiscent of the brute-force algorithm. 
Table \ref{table2} summarizes the proposed algorithm.

The complexity of the proposed algorithm---in terms of number of
times \eqref{gntilde} is solved---is 
\begin{equation}\textstyle
\big\lceil\log_2 (\frac{\mu_{\text{\sf max}} 
- \mu_{\text{\sf min}}}{\kappa})\big\rceil \times N(KM+2), 
\end{equation}
which is significantly less than the brute-force complexity 
in \eqref{complexitybrute}. 
Although the proposed algorithm is sub-optimal, the difference between 
the optimal GSRA utility $U_{\text{\sf GSRA}}^*$ and that attained
by the proposed algorithm 
$\hat{U}_{\text{\sf GSRA}}(\underline{\mu}, \bar{\mu})$, as 
$\underline{\mu} \to \bar{\mu}$, can be bounded as follows
\cite{aggarwal:TSP:2010}:
\begin{eqnarray}
U_{\text{\sf GSRA}}^* - \lim_{\underline{\mu} \to \bar{\mu}} 
\hat{U}_{\text{\sf GSRA}}(\underline{\mu}, \bar{\mu})
&\leqslant& (\mu^* - \mu_{\text{\sf min}})
\big(X\con - X_{\text{\sf tot}}^{t,\text{\sf pro}}(\mu^*)\big) \label{eq:DSRAbound} \\
&\leqslant& \Biggl\{\begin{array}{l@{~~}l} 0 & \textrm{if}~ |S_n(\mu^*)| \leqslant 1~\forall n \\
(\mu_{\text{\sf max}} - \mu_{\text{\sf min}})X\con
 & \textrm{otherwise} \end{array}. \label{eq:bound}
\end{eqnarray}
In \secref{simulations}, we evaluate \eqref{DSRAbound} by simulation, and show
that the performance loss is negligible.

\section{Updating the Posterior Distributions from ACK/NAK Feedback}	\label{sec:SSGupdate}

In this section, we propose a recursive procedure to compute the posterior
pdfs $p(\gamma_{n,k}^{t} \,|\, \vec{\mathbb{F}}_{-\infty}^{t-d})$
required by the proposed greedy algorithm in Table \ref{table2}
when the channel is first-order\footnote{
  The extension to higher-order Markov channels is straightforward.
} Markov.

Let the time-$t$ user-$k$ channel be described by the discrete-time channel impulse response $\vec{h}_{k}^{t} \defn [h_{1,k}^{t},\dots,h_{L,k}^{t}]^{\intercal} \in \mathbb{C}^L$, where 
$(\cdot)^\intercal$ denotes transpose. 
The corresponding frequency-domain subchannel gains 
$\vec{H}_{k}^{t} \defn [H_{1,k}^{t},\dots,H_{N,k}^{t}]^{\intercal} \in 
\mathbb{C}^N$ are then given by 
\begin{eqnarray}
\vec{H}_{k}^{t} &=& \vec{G}\vec{h}_{k}^{t}, \label{eq:G}
\end{eqnarray}
where the OFDMA modulation matrix $\vec{G}\in\Complex^{N\times L}$ contains
the first $L$ columns of the $N$-DFT matrix. 
Assuming additive white Gaussian noise with unit variance, the 
SSG of subchannel $n$ for user $k$ is given by
$\gamma_{n,k}^t = |H_{n,k}^{t}|^2$, and so we can write 
\begin{equation}
 p(\gamma_{n,k}^{t}\,|\,\vec{\mathbb{F}}_{-\infty}^{t-d}) = \int_{\vec{h}_{k}^{t}}  
p(\gamma_{n,k}^{t}\,|\,\vec{h}_{k}^{t}) p(\vec{h}_{k}^{t}\,|\,\vec{\mathbb{F}}_{-\infty}^{t-d})  \label{eq:gam|F}
\end{equation}
with $p(\gamma_{n,k}^{t}\,|\,\vec{h}_{k}^{t}) = \delta(\gamma_{n,k}^{t}-|\vec{e}_n^\intercal \vec{G}\vec{h}_k^t|^2)$, 
where $\delta(\cdot)$ is the
Dirac delta and $\vec{e}_n$ is the $n^{th}$ column of the identity matrix.
Using the channel's Markov property and Bayes rule, we find that
\begin{eqnarray}
  p(\vec{h}_{k}^{t}\,|\,\vec{\mathbb{F}}_{-\infty}^{t-d})
  &=& \int_{\vec{h}_{k}^{t-d}} 
   	p(\vec{h}_{k}^{t}\,|\,\vec{h}_{k}^{t-d}) \,
   	p(\vec{h}_{k}^{t-d}\,|\,\vec{\mathbb{F}}_{-\infty}^{t-d})
	\label{eq:hd|F} \\
  p(\vec{h}_{k}^{t-d}\,|\,\vec{\mathbb{F}}_{-\infty}^{t-d})
  &=& \frac{
   	p(\vec{f}_k^{t-d}\,|\,\vec{h}_{k}^{t-d}, \vec{\mathbb{F}}_{-\infty}^{t-d}
	\setminus \vec{f}_k^{t-d})\,p(\vec{h}_{k}^{t-d}\,|\,
	\vec{\mathbb{F}}_{-\infty}^{t-d}\setminus \vec{f}_k^{t-d})}
	{\int_{\vec{\bar{h}}_{k}^{t-d}}
   	p(\vec{f}_k^{t-d}\,|\,\vec{\bar{h}}_{k}^{t-d}, 
	\vec{\mathbb{F}}_{-\infty}^{t-d}
	\setminus \vec{f}_k^{t-d})\, p(\vec{\bar{h}}_{k}^{t-d}\,|\,
	\vec{\mathbb{F}}_{-\infty}^{t-d}\setminus \vec{f}_k^{t-d})}, \label{eq:condh}
\end{eqnarray}
where $\setminus$ denotes the set-difference operator.
Using the fact that $p(\vec{f}_k^{t-d}\,|\,
\vec{h}_{k}^{t-d}, \vec{\mathbb{F}}_{-\infty}^{t-d} \setminus \vec{f}_k^{t-d})
=p(\vec{f}_k^{t-d}\,|\,\vec{h}_{k}^{t-d}, \vec{I}^{t-d}, \vec{p}^{t-d})$, 
along with the fact that $(\vec{I}^{t-d}, \vec{p}^{t-d})$ is a 
deterministic function of $\vec{\mathbb{F}}_{-\infty}^{t-2d}$ (and therefore
of $\vec{\mathbb{F}}_{-\infty}^{t-d-1}$), we then have from \eqref{condh} that
\begin{eqnarray}
p(\vec{h}_{k}^{t-d}\,|\,\vec{\mathbb{F}}_{-\infty}^{t-d})
   &=& \frac{
   	p(\vec{f}_k^{t-d}\,|\,\vec{h}_{k}^{t-d}, \vec{I}^{t-d}, \vec{p}^{t-d})\,
   	p(\vec{h}_{k}^{t-d}\,|\,\vec{\mathbb{F}}_{-\infty}^{t-d-1})}
	{\int_{\vec{\bar{h}}_{k}^{t-d}}
   	p(\vec{f}_k^{t-d}\,|\,\vec{\bar{h}}_{k}^{t-d}, \vec{I}^{t-d}, \vec{p}^{t-d})
	\,p(\vec{\bar{h}}_{k}^{t-d}\,|\,\vec{\mathbb{F}}_{-\infty}^{t-d-1})}.
\label{eq:h|F2}
\end{eqnarray}
Using the Markov property again, we get
\begin{eqnarray}
  p(\vec{h}_{k}^{t-d}\,|\,\vec{\mathbb{F}}_{-\infty}^{t-d-1})
   &=& \int_{\vec{h}_{k}^{t-d-1}} p(\vec{h}_{k}^{t-d}\,|\,\vec{h}_{k}^{t-d-1}) \,
   	p(\vec{h}_{k}^{t-d-1}\,|\,\vec{\mathbb{F}}_{-\infty}^{t-d-1}).\label{eq:h1|F}
\end{eqnarray}

Recall that $f_{n,k}^t$, the feedback received about user $k$ on channel $n$ at time $t$, takes values from the set $\{0, 1, \emptyset\}$, where $0$ denotes a NAK, $1$ denotes an ACK, and $\emptyset$ denotes that user $k$ was not scheduled on subchannel $n$ at time $t$.
Assuming that, conditioned on $\vec{h}_k^t$, the feedbacks generated by user $k$ are independent across subchannels, we have
\begin{align}
  p(\vec{f}_{k}^{t}|\vec{h}_{k}^{t}, \vec{I}^{t}, \vec{p}^{t})
  &= \prod_{n=1}^N p(f_{n,k}^{t}|\vec{h}_{k}^{t}, \vec{I}^{t}, 
	\vec{p}^{t}), \label{eq:f|h} \\
  p(f_{n,k}^{t} = f\,|\, \vec{h}_{k}^{t}, \vec{I}^{t}, \vec{p}^{t})
  &= \begin{cases}
	\sum_m I_{n,k,m}^{t}a_m e^{-b_m p_{n,k,m}^{t}\gamma_{n,k}^{t}} &\textrm{if}~ f=0\\
	\sum_m I_{n,k,m}^{t}\left(1-a_me^{-b_m p_{n,k,m}^{t}\gamma_{n,k}^{t}}\right) 
	&\textrm{if}~ f=1\\
   	1-\sum_m I_{n,k,m}^{t} &\textrm{if}~ f=\emptyset, \label{eq:f|gam}
   	\end{cases}
\end{align}
where $\gamma_{n,k}^t = |H_{n,k}^t|^2$ can be determined from 
$\vec{h}_{k}^{t}$ via \eqref{G}.
Together, \eqref{gam|F}-\eqref{f|gam} suggest a method of 
\emph{recursively} updating the channel distributions,
using the new feedback obtained at each time $t$, 
which is given in \tabref{snr}.

We now propose the use of particle filtering \cite{pf2}
to circumvent the evaluation of multidimensional integrals 
in the recursion of \tabref{snr}. 
Particle filtering is a well-known technique that approximates the pdf 
of a random variable using a suitably chosen probability mass function (pmf).
In the sequel, for simplicity of illustrations, we assume a 
Gauss-Markov model of the form
\begin{equation}
h_{l,k}^{t+1} \;=\; (1-\alpha) h_{l,k}^{t} + \alpha w_{l,k}^{t}, \label{eq:channel}
\end{equation}
where $w_{l,k}^{t}$ is unit-variance circular Gaussian 
and $\alpha \in (0, 1]$ is a known constant that determines the fading rate. 
Here, $w_{l,k}^{t}$ is assumed to be i.i.d. for all $t,l,k$. 
At each time-step $t$, for $k \in \{1, \ldots K\}$, we use $S$ 
particles in the approximations
\begin{eqnarray}
p(\vec{h}_{k}^t \,|\, \vec{\mathbb{F}}_{-\infty}^{t-d}) 
&\approx& \sum_{i=1}^S \nu_{k}^{t\,|\,t-d}[i] \, 
	\delta(\vec{h}_{k}^t - \vec{h}_{k}^t[i]), 
~\textrm{and} \nonumber \\
p(\vec{h}_{k}^{t-d} \,|\, \vec{\mathbb{F}}_{-\infty}^{t-d}) 
&\approx& \sum_{i=1}^S \nu_k^{t-d\,|\,t-d}[i] \, 
	\delta(\vec{h}_{k}^{t-d} - \vec{h}_{k}^{t-d}[i]), 
\label{eq:particle}
\end{eqnarray}
where 
$\vec{h}_k^t[i]=\big[h_{1,k}^t[i], \ldots, h_{L,k}^t[i]\big]^\intercal \in \mathbb{C}^L$
denotes the $i^{\textrm{th}}$ (vector) particle, for $i \in \{1, \ldots, S\}$,
and $\nu_{k}^{t_1|t_2}[i] \in \mathbb{R}^+$
is the probability mass assigned to the particle 
$\vec{h}_k^{t_1}[i]$ based on the observations received up to time $t_2$.
The steps to recursively compute these particles and their corresponding weights
are detailed in \tabref{particle}. 

Using the approximation in \eqref{particle}, we note that the expectation of
any function of subchannel-gain, $\vec{h}_k^t$, can be found using
\begin{equation}
\E\{A(\vec{h}_k^t)\} \approx \sum_{i}
\nu_{k}^{t\,|\, t-d}[i] A\big(\vec{h}_k^{t}[i]\big),
\end{equation}
where $A(\cdot)$ is an arbitrary function. 
Recalling that the SSG $\gamma_{n,k}^t$ is a deterministic 
function of the subchannel-gain, $\vec{h}_k^t$, 
any function of $\gamma_{n,k}^t$ is also a function of $\vec{h}_k^t$.

\section{Numerical Results} \label{sec:simulations}

In this section, we numerically evaluate the performance of the 
proposed greedy scheduling and resource allocation from
\secref{greedy} with the posterior update from \secref{SSGupdate}.
For this, we consider an OFDMA system with independent 
first-order Gauss-Markov channels \eqref{channel}. 
We assumed, if not otherwise stated, $K=8$ available users, $N=32$ OFDMA 
subchannels, channel fading parameter $\alpha = 10^{-3}$ and impulse 
response length $L=2$. 
We used the modulation matrix 
$\vec{G} = \sqrt{\beta} \vec{F} \in \mathbb{C}^{N \times L}$ (recall 
\eqref{G}), where $\vec{F}$ contains the first $L(\leqslant N )$ columns of the 
unitary $N$-DFT matrix and $\beta = \frac{N}{L} \frac{2-\alpha}{\alpha}$
ensures that the variance of $H_{n,k}^t$ is unity for all $(n,k)$. 
Thus, the mean of the SSG $\gamma_{n,k}^t$ was also unity for all $(n,k)$.
Since the subchannel-averaged total transmit power equals 
$\frac{1}{N}\sum_{n,k,m} I_{n,k,m}^t P_{n,k,m}^t = 
\frac{1}{N}\sum_{n,k,m} X_{n,k,m}^t = X\con/N$, it is readily seen that the
average per-subchannel signal-to-noise ratio is 
$\textsf{SNR}\defn \E\{\frac{1}{N}\sum_{n,k,m} X_{n,k,m}^t\gamma_{n,k}^t\}
= \frac{1}{N}\sum_{n,k,m} X_{n,k,m}^t\E\{\gamma_{n,k}^t\}
= X\con/N$.
For the plots, we averaged $500$ realizations, each with
$100$ time-slots. 
Of these $100$ time-slots, the first $50$ were ignored to avoid 
transient effects. 

For illustrative purposes, we assumed uncoded $2^{m+1}$-QAM signaling 
with MCS index $m \in \{1, \ldots, 15\}$. 
In this case, we have $r_m=m+1$ bits per symbol, one symbol per ``codeword,''
and one codeword per packet. 
In the packet error-rate model $\epsilon = a_m e^{-b_m P \gamma}$, we 
assumed $a_m = 1$ and $b_m = 1.5/(2^{m+1}-1)$ because the symbol 
error-rate of a $2^{m+1}$-QAM system is well approximated by 
$\exp(-1.5 P \gamma/(2^{m+1}-1))$ in the high-$(P \gamma)$ regime 
\cite{Proakis:Book:08} and is $\approx 1$ when $P \gamma= 0$. 
Throughout, we used the identity utility (i.e., $U_{n,k,m}(x) = x$ 
for all $n,k,m$) so that the objective was maximization of sum goodput,
and we assumed a feedback delay of $d=1$. 

The performance of the proposed greedy algorithm was compared to three 
reference schemes: fixed-power random user scheduling (FP-RUS), the 
``causal global genie'' (CGG), and the ``non-causal global genie'' (NCGG). 
The FP-RUS scheme schedules users uniformly at random, allocates
power uniformly across subchannels, and selects the MCS to maximize
expected goodput.
The FP-RUS, which makes no use of feedback, should perform no better
than any feedback-based scheme.
The CGG (recall \secref{causalgenie}) performs optimal 
scheduling and resource allocation under perfect knowledge of all 
SSGs at the previous time-instant (since $d=1$), i.e., given
$\{\gamma_{n,k}^{t-1}~ \forall n,k\}$ at time $t$. 
From \lemref{genie_bnd}, we know that the CGG upper-bounds the POMDP.
The NCGG is similar to the CGG, but assumes perfect 
knowledge of all SSGs at \emph{all} times, i.e., 
given $\{\gamma_{n,k}^{\tau}~ \forall n,k,\tau\}$ at time $t$.
Thus, it provides an upper bound on the CGG that is invariant to 
fading rate $\alpha$. 
The NCGG has a greedy implementation, like the CGG, 
but without the conditional expectation in \eqref{CGGeq}. 

\Figref{goodput_1e-3} shows a typical realization of instantaneous 
sum-goodput versus time $t$, when $\alpha = 10^{-3}$. 
There, one can see a large gap between the FP-RUS and the CGG, and a 
much smaller gap between the CGG and the NCGG.
The proposed scheme starts without CSI, and initially performs no 
better than the FP-RUS.
From ACK/NAK feedbacks, however, it quickly learns the CSI 
well enough to perform scheduling and resource allocation at a level 
that yields sum-goodput much closer to the CGG than to the FP-RUS.

\Figref{versus_Np} plots average sum-goodput versus
the number of particles $S$ used to update the posterior 
distributions in the proposed greedy scheme (recall \secref{SSGupdate}). 
There we see that the performance of the proposed scheme increases with $S$, 
but shows little improvement for $S>30$. 
Thus, $S=30$ particles were used to construct the other plots.
Remarkably, with only $S=5$ particles,
the proposed algorithm captures a significant portion of the 
maximum possible goodput gain over the FP-RUS.

\Figref{versus_alp} plots average sum-goodput versus the fading rate $\alpha$.
There we see that, at low fading rates (i.e., small $\alpha$), the 
proposed greedy scheme achieves an average sum-goodput that is 
much higher than the FP-RUS and, in fact, not far from the CGG upper bound.
For instance, at $\alpha=10^{-4}$, the sum-goodput attained by the proposed 
scheme is $92$\% of the upper bound and $170$\% of that attained by the FP-RUS.
As the fading rate $\alpha$ increases, we see that the sum-goodput attained
by the proposed scheme decreases, and eventually converges to that of the 
FP-RUS.
This behavior is due to the fact that, as $\alpha$ increases, it becomes 
more difficult to predict the SSGs using delayed ACK/NAK feedback, thereby
compromising the scheduling-and-resource-allocation decisions that are
made based on the predicted SSGs.
In fact, one can even observe a gap between the CGG and NCGG for large 
$\alpha$ because, even with delayed perfect-SSG feedback, the current SSGs 
are difficult to predict.

\Figref{versus_alp} reveals a gap between the proposed scheme and the CGG 
bound that persists as $\alpha\rightarrow 0$.
This non-vanishing gap can be attributed---at least in part---to 
\emph{greedy} scheduling under ACK/NAK feedback. 
Intuitively, we have the following explanation. 
Because the inferred SSG-distributions of 
not-recently-scheduled users quickly revert to their apriori 
form, the proposed greedy algorithm 
will continue to schedule users as long as their SSGs remain 
better than the apriori value. 
There may exist, however, not-recently-scheduled users with far better 
SSGs who remain invisible to the proposed scheme, only because they have not
recently been scheduled.

Figures~\ref{fig:versus_N_varying_Pcon}~and~\ref{fig:versus_N} plot average 
sum-goodput versus the number of subchannels (i.e., total bandwidth) $N$. 
In \figref{versus_N_varying_Pcon}, the total BS power $X\con$ is scaled with 
$N$ such that the per-subchannel \textsf{SNR} 
remains fixed at $10$dB, whereas, in \figref{versus_N}, 
the total BS power $X\con$ remains invariant to the bandwidth $N$, and is 
set such that per-subchannel $\textsf{SNR} = 10$dB for $N=32$.
In both cases, the average sum-goodput increases with bandwidth $N$, 
as expected, since the availability of more subchannels increases 
not only scheduling flexibility, but also the possibility of stronger 
subchannels, which can be exploited by the BS. 
In \figref{versus_N_varying_Pcon}, where the per-subchannel \textsf{SNR} 
is fixed, the sum-goodput increases linearly with bandwidth $N$, 
as expected.
In all cases, the proposed greedy scheme captures about $80\%$
of the sum-goodput gain achievable over the FP-RUS.

\Figref{versus_K} plots average sum-goodput versus the number of available 
users $K$.
It shows that, as $K$ increases, the average sum-goodputs achieved by 
the NCGG, CGG, and the proposed greedy schemes increase, whereas 
that achieved by the FP-RUS remains constant. 
This behavior results because, with the former 
schemes, the availability of more users can be exploited 
to schedule users with stronger subchannels, whereas with the 
FP-RUS scheme, this advantage is lost due to the complete lack of information 
about the users' instantaneous channel conditions. 
\Figref{versus_K} also suggests that, as $K$ increases, 
the sum-goodput of the proposed greedy scheme saturates.
This can be attributed to the fact that the proposed greedy 
algorithm can only track the channels of recently scheduled users,
and thus cannot benefit directly from the growing pool of 
not-recently-scheduled users.

In \Figref{versus_SNR}, the top subplot shows average sum-goodput 
versus \textsf{SNR},
while the bottom subplot shows the average value of the bound 
\eqref{DSRAbound} on the optimality gap of our proposed approach 
to the GSRA problem, also versus \textsf{SNR}. 
The top plot shows that, as the \textsf{SNR} increases, the 
proposed greedy scheme continues to perform much closer to the 
NCGG/CGG bounds than it does to the FP-RUS scheme. 
The bottom plot establishes that the sum-goodput loss due to the 
sub-optimality in the algorithm used to attack the GSRA problem
is negligible, e.g., at most $0.0025$\% over all \textsf{SNR}.

\section{Conclusion}				\label{sec:conclusion}

In this paper, we considered the problem of joint scheduling and resource 
allocation in the OFDMA downlink under ACK/NAK feedback, with the goal of 
maximizing an expected long-term goodput-based utility subject to an
instantaneous sum-power constraint. 
First, we established that the optimal solution to the problem is a partially 
observable Markov decision process (POMDP), which is impractical to implement.
Consequently, we proposed a greedy approach to joint scheduling and resource 
allocation based on the posterior distributions of the squared subchannel gain 
(SSG) for every user/subchannel pair, which has polynomial complexity.
Next, for Markov channels, we outlined a recursive method to update the 
posterior SSG distributions from the ACK/NAK feedbacks received at each 
time-slot, and proposed an efficient implementation based on 
particle filtering. 
To gauge the performance of our greedy scheme relative to that of the 
optimal POMDP (which is impossible to implement), we derived a performance
upper-bound on POMDP, known as the causal global genie (CGG). 
Numerical experiments suggest that our greedy scheme achieves a significant 
fraction of the maximum possible performance gain over fixed-power random 
user scheduling (FP-RUS), despite its low-complexity implementation.
For example, a representative simulation 
using 
$N=32$ OFDMA subchannels, 
$K=8$ available users,
\textsf{SNR}$=10$dB, 
and $S=30$ particles,
shows that the sum-goodput of the proposed scheme is $92$\%
of the upper bound and $170$\% of that attained by the FP-RUS 
(see \figref{versus_alp}).


\bibliographystyle{ieeetr}
\bibliography{macros_abbrev,references}
\def\baselinestretch{1.0}

\begin{table}[h]\footnotesize
\caption{Brute-force steps for a given $\vec{I}$}
\begin{enumerate}
\item Initialize $\underline{\mu} = \mu_{\text{\sf min}}$ and 
	$\bar{\mu} = \mu_{\text{\sf max}}$. 
  \item Set 
	$\mu = \frac{\underline{\mu} + \bar{\mu}}{2}$.
\item For each $(n,k,m)$,
	\begin{enumerate}
	\item Use \eqref{P_n^*}-\eqref{gntildeb} to obtain $P_{n,k,m}^{*}(\mu)$.
	\end{enumerate}
\item Calculate $X\overall^{*}(\vec{I},\mu) \defn \sum_{n,k,m}I_{n,k,m}\, P_{n,k,m}^{*}(\mu)$. 
\item If $X\overall^{*}(\vec{I},\mu) > X\con\,$, 
set $\underline{\mu} = \mu$, otherwise set $\bar{\mu} = \mu$.
\item If $\bar{\mu} - \underline{\mu} > \kappa$, go to step 2), else proceed to step 7).
\item If $X\overall^*(\vec{I},\bar{\mu}) \neq X\overall^*(\vec{I},\underline{\mu})$, set 
$\lambda = \frac{X\overall^*(\vec{I},\underline{\mu}) - 
X\con}{X\overall^*(\vec{I},\underline{\mu}) - 
X\overall^*(\vec{I},\bar{\mu})}$, otherwise set $\lambda = 0$. 
\item Set $\hat{\mu}_{\vec{I}} = \bar{\mu}$. 
The best power allocation is given by 
$\hvec{P}(\vec{I}) = \lambda \vec{P}^{*}(\bar{\mu})
+ (1-\lambda) \vec{P}^{*}(\underline{\mu})$, and
$\hat{L}_{\vec{I}} = L^t_{\vec{I}}(\bar{\mu}, \hvec{P}(\vec{I}))$ 
gives the best Lagrangian value.
\end{enumerate}
\label{table1}
\end{table}

\begin{table}[ht]\footnotesize
\caption{Proposed greedy algorithm}
\begin{enumerate}
  \item Initialize 
  	$\underline{\mu} = \mu_{\text{\sf min}}$ and 
  	$\bar{\mu} = \mu_{\text{\sf max}}$. 
  \item Set 
  	$\mu = \frac{\underline{\mu}+\bar{\mu}}{2}$.
   \item For each subchannel $n=1,\dots,N$:
   \begin{enumerate}
    \item For each $(k,m)$,
    \begin{enumerate}
     \item Use \eqref{P_n^t*}-\eqref{gntilde} to calculate $P_{n,k,m}^{t,*}(\mu)$.
     \item Use $P_{n,k,m}^{t,*}(\mu)$ to calculate 
	$V_{n,k,m}^t(\mu, P_{n,k,m}^{t,*}(\mu))$ via \eqref{V}.
    \end{enumerate}
   \item Calculate $S_n^t(\mu)$ using \eqref{Snmu}.
   \end{enumerate}
   \item Find $\vec{I}^{t,\text{\sf pro}}(\mu)$ using \eqref{I*}.
   \item Calculate $X\overall^{t,\text{\sf pro}}(\mu) = \sum_{n,k,m} I^{t,\text{\sf pro}}_{n,k,m}(\mu) \, 
	P^{t,*}_{n,k,m}(\mu)$.
   \item If $X\overall^{t,\text{\sf pro}}(\mu) > X\con$, 
	set $\underline{\mu} = \mu$, 
   otherwise set $\bar{\mu} = \mu$.
   \item If $\bar{\mu} - \underline{\mu} > \kappa$, go to 
   step 2), else proceed to step 8).
 \item Now we have $\mu^* \in [\underline{\mu}, 
 \bar{\mu}]$ and $\bar{\mu} - \underline{\mu} < \kappa$. 
For both $\vec{I} = \vec{I}^{t,\text{\sf pro}}(\underline{\mu})$ and $\vec{I}=\vec{I}^{t,\text{\sf pro}}(\bar{\mu})$ (since they may differ),
  calculate $\hvec{P}(\vec{I})$ and 
	$\hat{L}_{\vec{I}}$ as described for the brute force algorithm.
  \item Choose $\hvec{I}^t
	= \argmin_{\vec{I} \in \{\vec{I}^{t,\text{\sf pro}}(\underline{\mu}),\, 
	\vec{I}^{t,\text{\sf pro}}(\bar{\mu})\}} \hat{L}_{\vec{I}}$ as the user-MCS allocation 
	and $\hvec{P}^t = \hvec{P}(\hvec{I}^t)$ as the associated 
	power allocation.
\end{enumerate}
\label{table2}
\end{table}

\begin{table}[ht]\footnotesize
\caption{Recursive update of channel posteriors}
At time $t$, for each user $k$, the pdf $p(\vec{h}_{k}^{t-d-1}\,|\, \vec{\mathbb{F}}_{-\infty}^{t-d-1})$ is available from the previous time-instant. The user-$k$ recursion is then
    \begin{enumerate}
      \item
    	Observe new feedbacks $\vec{f}_{k}^{t-d}\in\{0,1,\emptyset\}^N$.
      \item Compute $p(\vec{h}_{k}^{t-d}\,|\,\vec{\mathbb{F}}_{-\infty}^{t-d-1})$
	using \eqref{h1|F}.
      \item
	Compute $p(\vec{f}_{k}^{t-d}\,|\,\vec{h}_{k}^{t-d}, \vec{I}^{t-d}, 
	\vec{P}^{t-d})$ using the error-rate rule \eqref{f|h}-\eqref{f|gam}.
      \item
	Using the distributions obtained in steps 2) and 3), compute 
	$p(\vec{h}_{k}^{t-d}\,|\, \vec{\mathbb{F}}_{-\infty}^{t-d})$ 
	via Bayes-rule step in \eqref{h|F2}.
      \item
	Compute $p(\vec{h}_{k}^{t}\,|\,\vec{\mathbb{F}}_{-\infty}^{t-d})$ 
	using the Markov-prediction step \eqref{hd|F}.
      \item
	For each $n$, compute $p(\gamma_{n,k}^t\,|\,
	\vec{\mathbb{F}}_{-\infty}^{t-d})$ via \eqref{gam|F}.
    \end{enumerate}
\label{tab:snr}
\end{table}

\begin{table}[ht]\footnotesize
\caption{Particle filtering steps}
Let the system begin at time-instant $t_0$. If $t \in \{t_0, \ldots, t_0 + d-1\}$:
\begin{enumerate}
\item Initialize $\{h_{l,k}^{t}[i] ~\forall i, k, l\}$
by drawing i.i.d. samples from $\mathcal{CN}(0, \frac{\alpha}{2-\alpha})$.
\item Set the importance weights $\nu_{k}^{t|t}[i] = \frac{1}{S} ~\forall k,i$.
\end{enumerate}
For any other time-instant $t~(\geqslant t_0 + d)$:
\begin{enumerate}
\item Using the previous samples $\{h_{l,k}^{\tau}[i] ~\forall l,k,i, \tau : \tau \leqslant t-d\}$,
obtain new samples according to the underlying Markov model as follows
\begin{equation}
h_{l,k}^{t}[i] = (1-\alpha)^d\,h_{l,k}^{t-d}[i] + \alpha\sum_{j=0}^{d-1} (1-\alpha)^j\, y_{l,k}^{t-j}[i], ~\forall i, l, k,
\nonumber
\end{equation}
where $y_{l,k}^{t-j}[i]$ is drawn i.i.d from $\mathcal{CN}(0,1)$ for all $i,l,k,j$.

\item For each user $k$,
\begin{enumerate}

\item Using the received feedbacks $\vec{f}_k^{t-d}$ and the set 
of importance weights from time $(t-d-1)$, i.e., $\{\nu_{k}^{t-d-1|t-d-1}[i]~\forall i\}$, compute the new set of importance weights at time $(t-d)$ using 
\begin{equation*}
\nu_{k}^{t-d|t-d}[i] = \nu_{k}^{t-d-1|t-d-1}[i] \times 
p\big(\vec{f}_k^{t-d}\,\big|\, \vec{h}_k^{t-d} = \vec{h}_{k}^{t-d}[i], \vec{I}^{t-d}, \vec{P}^{t-d}\big), 
\end{equation*}
for all $i$, where $p\big(\vec{f}_k^{t-d}\,\big|\, \vec{h}_k^{t-d} = \vec{h}_{k}^{t-d}[i], 
\vec{I}^{t-d}, \vec{P}^{t-d}\big)$ is given by \eqref{f|h}-\eqref{f|gam}.

\item Normalize the weights via
\begin{equation*}
\nu_{k}^{t-d|t-d}[i] \leftarrow \frac{\nu_{k}^{t-d|t-d}[i]}{
\sum_j \nu_{k}^{t-d|t-d}[j]} ~\forall i.
\end{equation*}

\item Compute the weights for the posterior distribution, 
$p(\vec{h}_{k}^t \,|\, \vec{\mathbb{F}}_{-\infty}^{t-d})$ using
\begin{equation*}
\nu_{k}^{t| t-d}[i]
= \sum_{j=1}^S \nu_{k}^{t-d|t-d}[j] \: p(\vec{h}_{k}^t = \vec{h}_{k}^{t}[i]\,
\big|\, \vec{h}_{k}^{t-d} = \vec{h}_{k}^{t-d}[j]).
\end{equation*}

\item Normalize the weights via
\begin{equation*}
\nu_{k}^{t|t-d}[i] \leftarrow \frac{\nu_{k}^{t|t-d}[i]}{
\sum_j \nu_{k}^{t|t-d}[j]} ~\forall i.
\end{equation*}

\end{enumerate}
\end{enumerate}
\label{tab:particle}
\end{table}

\clearpage

\putFrag{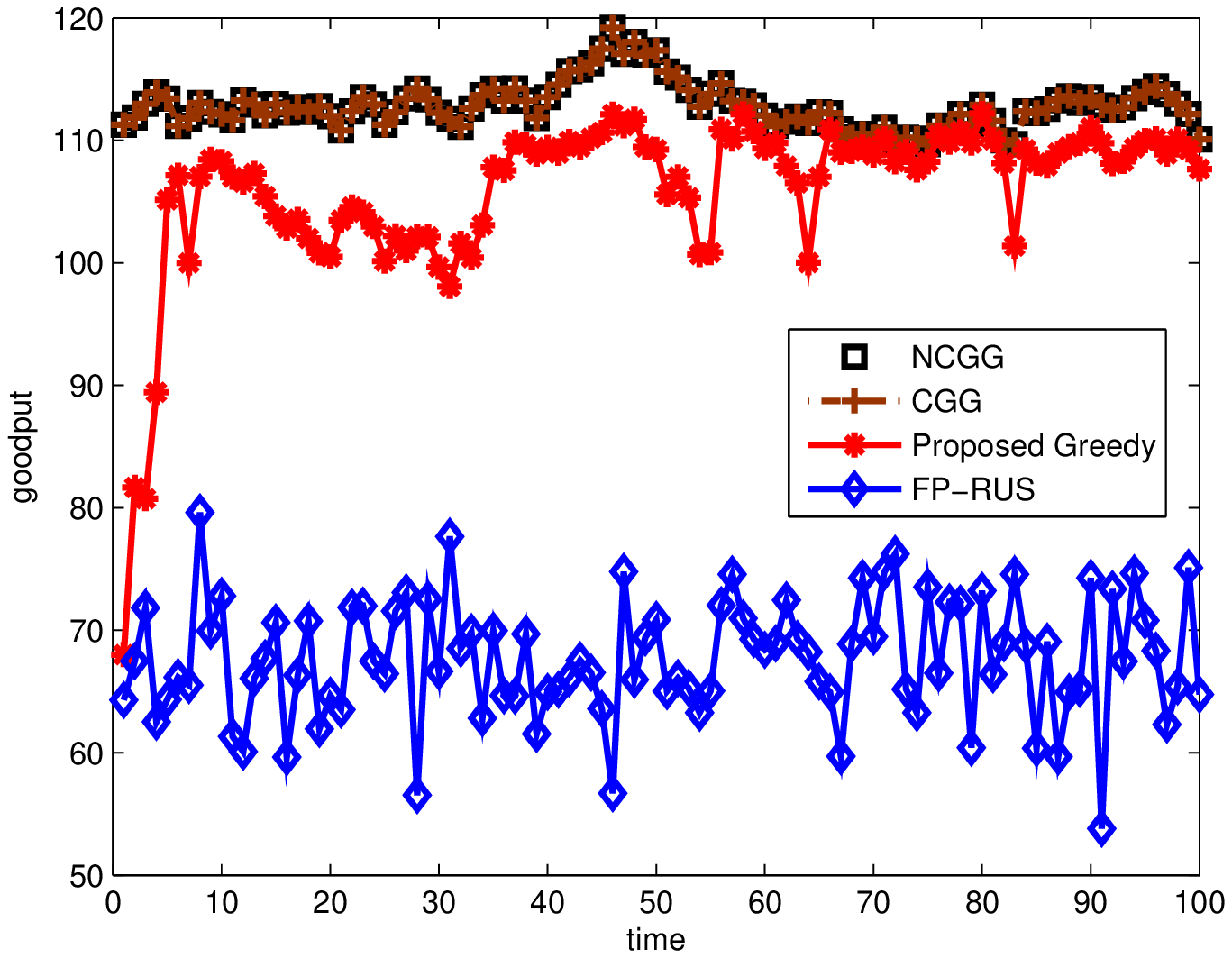}
	{Typical instantaneous sum-goodput versus time $t$. 
	 Here, $N=32$, $K=8$, $\textsf{SNR} = 10$dB, $\alpha = 10^{-3}$, 
	 and $S=30$.}
	{\figsize}
	{\psfrag{Packet Number}[][][0.7]{\sf Packet index}
	\psfrag{sum goodput}[][][0.7]{\sf Instantaneous sum-goodput (bpcu)}
	}

\putFrag{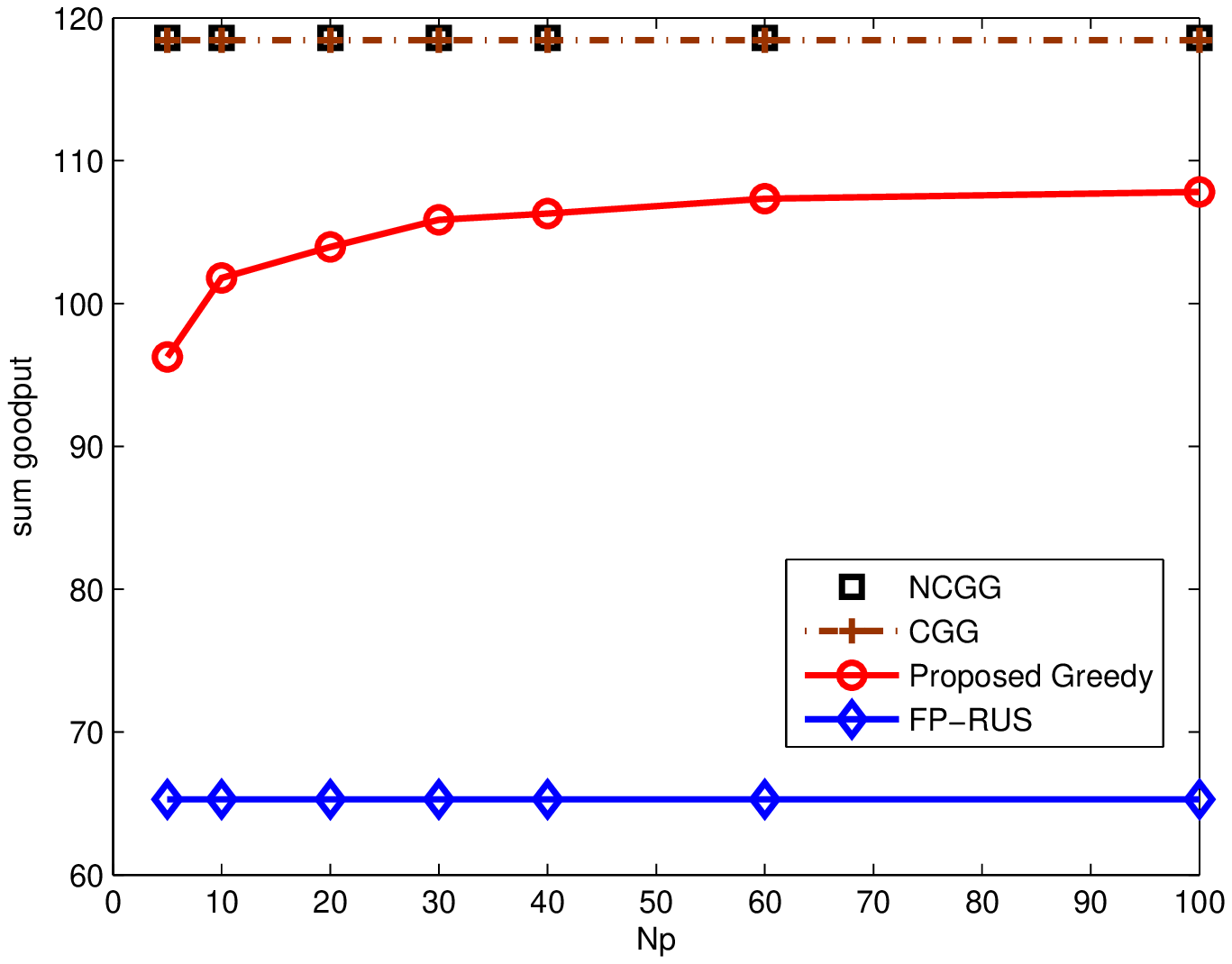}
	{Average sum-goodput versus the number of particles used to update 
	the channel posteriors. Here, $N=32$, $K=8$, $\textsf{SNR} = 10$dB, 
	and $\alpha = 10^{-3}$.}
	{\figsize}
	{\psfrag{Np}[][][0.7]{\sf number of particles}
	\psfrag{sum goodput}[][][0.7]{\sf Average sum-goodput (bpcu)}
	}

\putFrag{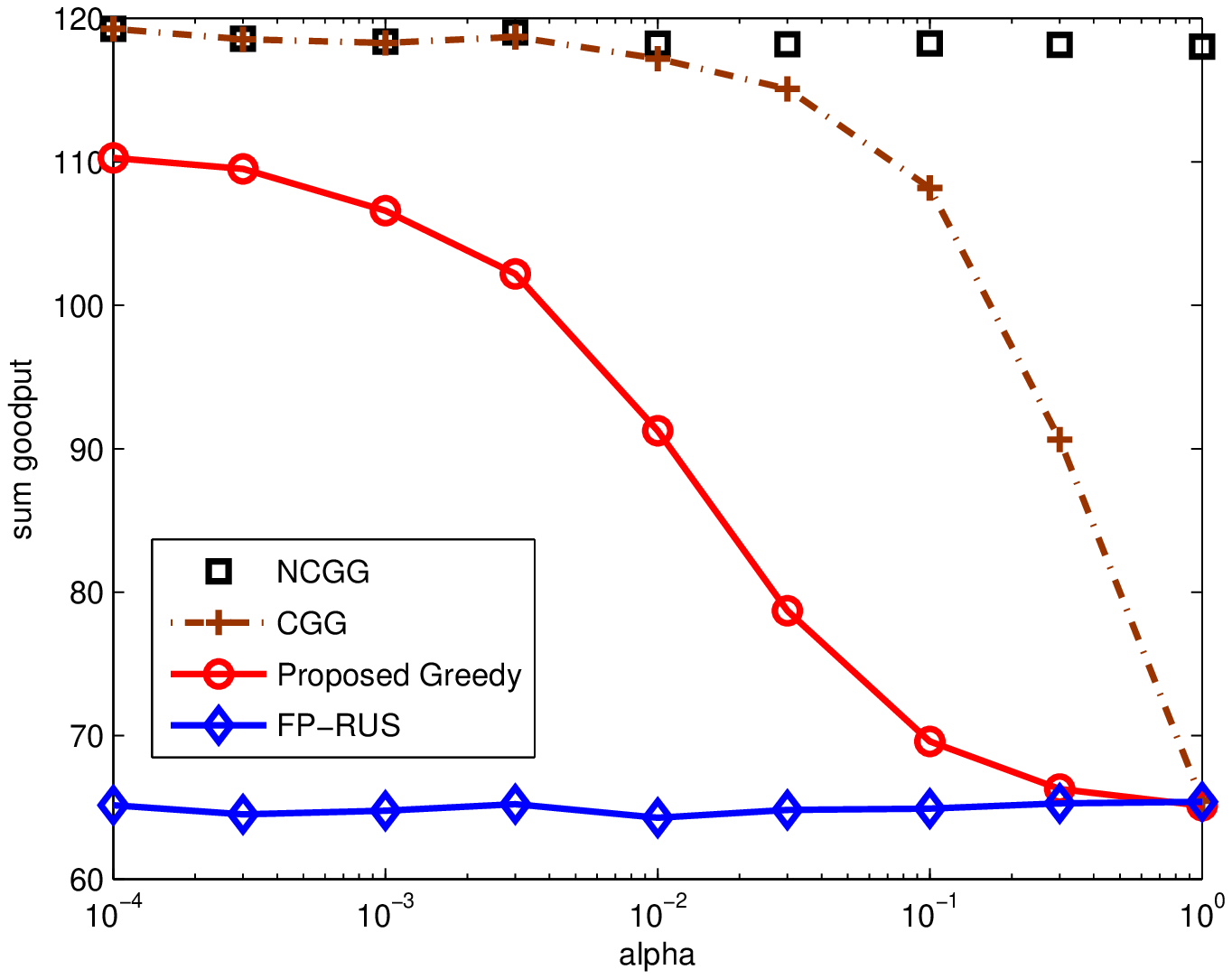}
	{Average sum-goodput versus fading rate $\alpha$. Here, $N=32$, 
	$K=8$, $\textsf{SNR} = 10$dB, and $S=30$.}
	{\figsize}
	{\psfrag{alpha}[][][0.7]{\sf Fading rate $\alpha$}
	\psfrag{sum goodput}[][][0.7]{\sf Average sum-goodput (bpcu)}
	}

\putFrag{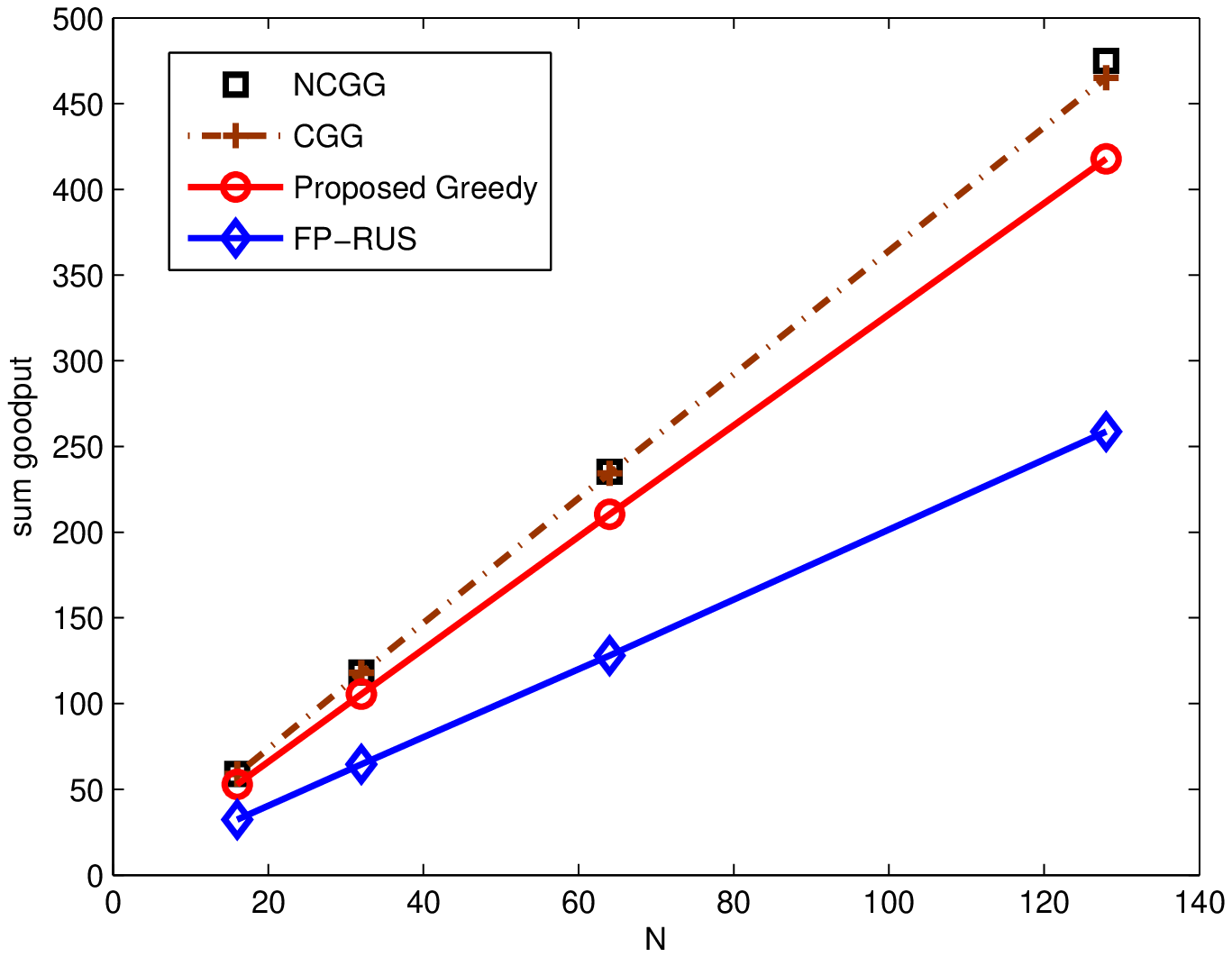}
	{Average sum-goodput versus number of subchannels $N$. 
	Here, $K = 8$, $\textsf{SNR} = 10$dB, 
	$\alpha = 10^{-3}$, and $S=30$.}
	{\figsize}
	{\psfrag{subchannels}[][][0.7]{\sf Number of subchannels, $N$}
	\psfrag{sum goodput}[][][0.7]{\sf Average sum-goodput (bpcu)}
	}

\putFrag{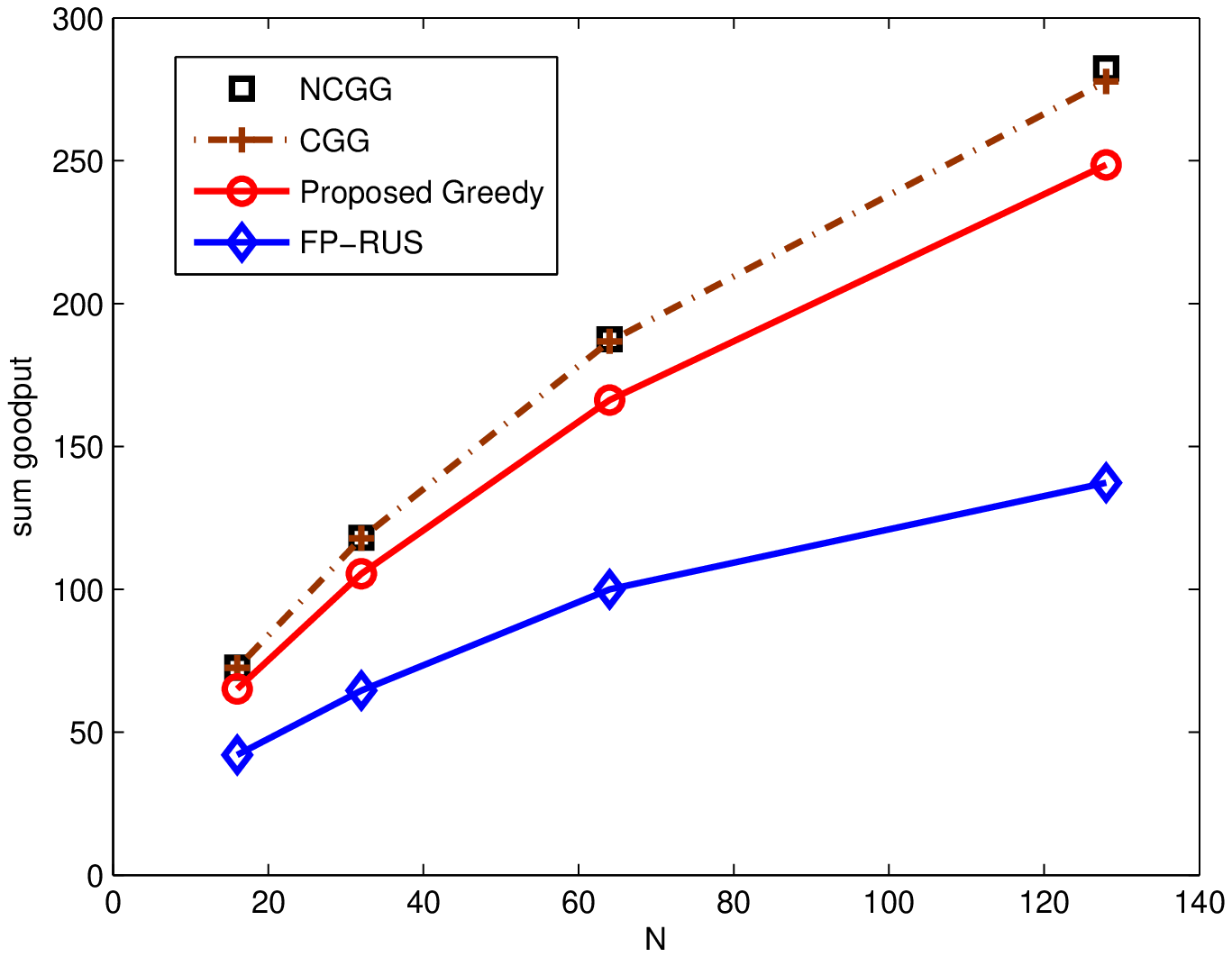}
	{Average sum-goodput versus number of subchannels $N$. 
	Here, $K = 8$, $X\con$ does not scale with $N$ and it is chosen such 
	that $\textsf{SNR} = 10$dB for $N=32$, $\alpha = 10^{-3}$, and 
	$S=30$.}
	{\figsize}
	{\psfrag{subchannels}[][][0.7]{\sf Number of subchannels, $N$}
	\psfrag{sum goodput}[][][0.7]{\sf Average sum-goodput (bpcu)}
	}

\putFrag{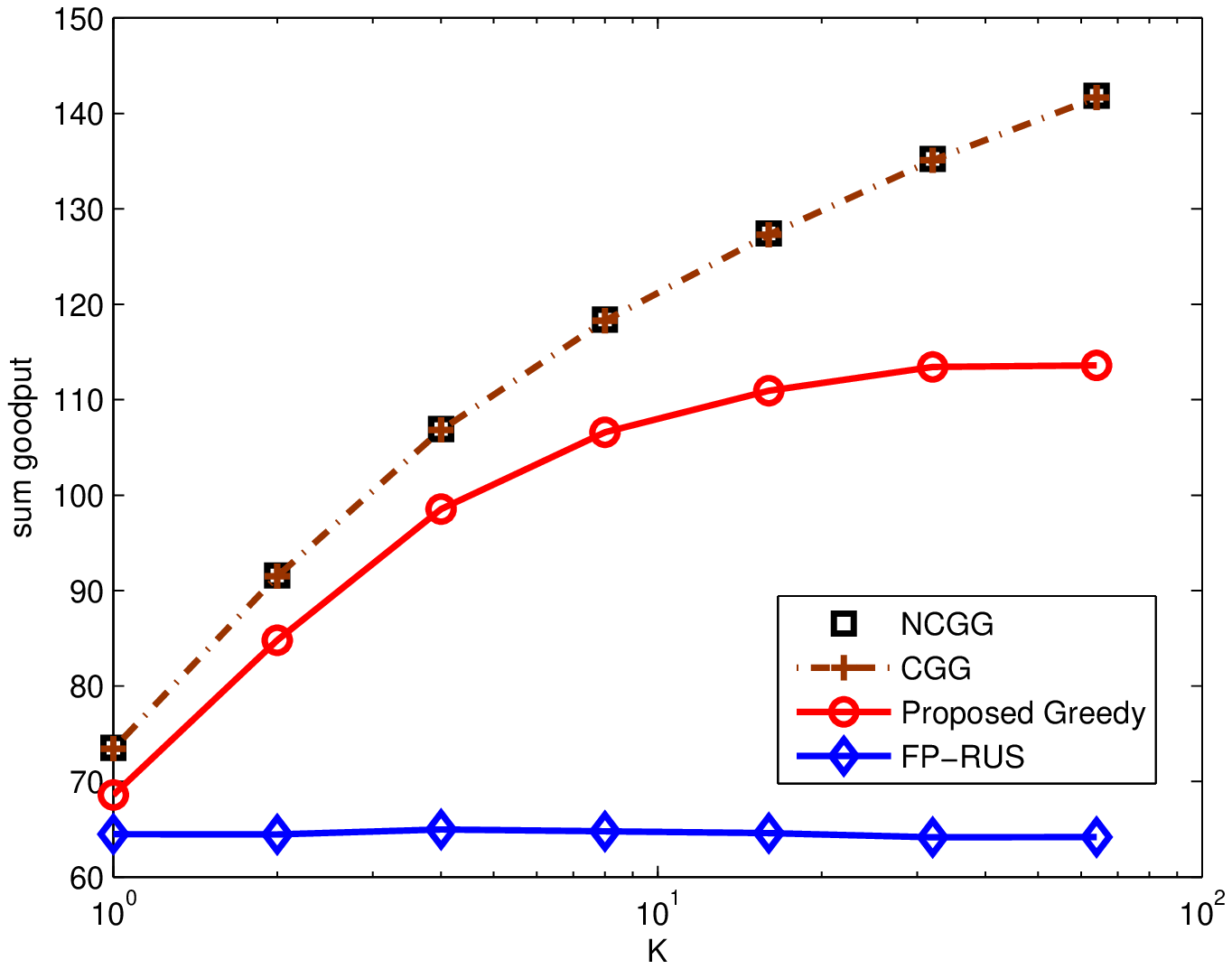}
	{Average sum-goodput versus number of users. In this plot, $N = 32$, 
	$\textsf{SNR} = 10$dB, $\alpha = 10^{-3}$, and $S=30$.}
	{\figsize}
	{\psfrag{K}[][][0.7]{\sf Number of users, $K$}
	\psfrag{sum goodput}[][][0.7]{\sf Average sum-goodput (bpcu)}
	}

\putFrag{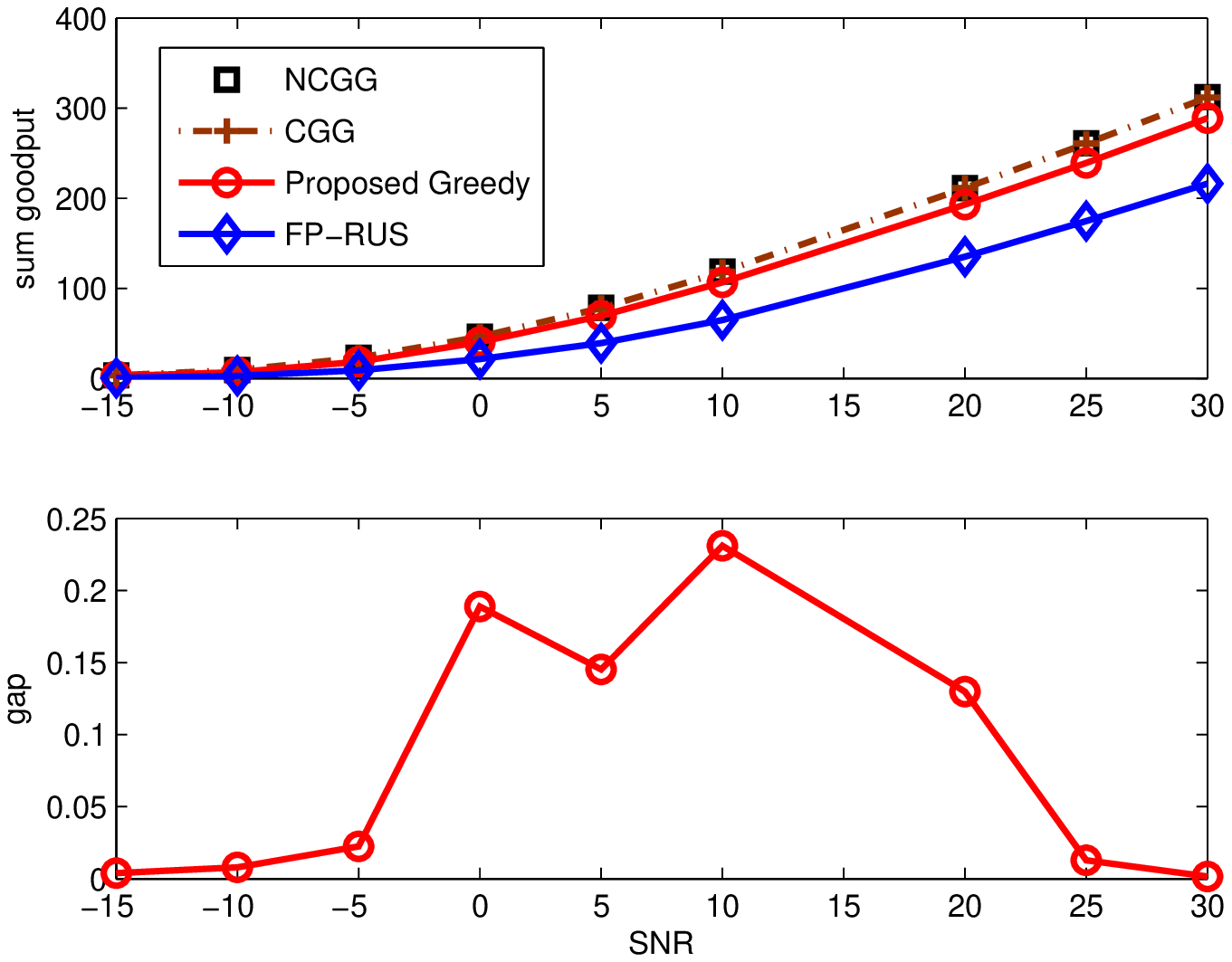}
	{The top plot shows the average sum-goodput as 
	a function of $\textsf{SNR}$. The bottom plot shows the average bound 
	on the optimality gap between the proposed and optimal
	greedy solutions (given in \eqref{DSRAbound}), i.e., the average 
	value of $(\mu^* - \mu_{\text{min}}) (X\con - 
	X\overall^*(\vec{I}^{\scriptscriptstyle\text{min}}, \mu^*))$. In 
	this plot, $N=32$, $K=8$, $\alpha = 10^{-3}$, and $S=30$.}
	{\figsize}
	{\psfrag{SNR}[][][0.7]{\textsf{SNR (in dB)}}
	\psfrag{sum goodput}[][][0.7]{$\begin{array}{c} 
	\textsf{Average sum-goodput}  \\ \textsf{(bpcu)} \\ \\ 
	\end{array}$}
	\psfrag{gap}[][][0.7]
	{$\begin{array}{c} \textsf{Optimality gap in the} \\ \textsf{average sum-goodput} \\
	\textsf{(bpcu)}
        \\ \\ \\ \end{array} \vspace{0.5cm}$}}

\end{document}